\documentclass[aps,prc,preprint,showpacs,preprintnumbers,nofootinbib,float,superscriptaddress,longbibliography]{revtex4-1}
\usepackage{graphicx, fancybox}
\usepackage{amsmath,amssymb}
\usepackage[colorlinks=true, pdfstartview=FitV, linkcolor=red,
		     citecolor=blue, urlcolor=blue]{hyperref}
\usepackage{color}
\usepackage{soul}
\usepackage{url}
\usepackage[normalem]{ulem}
\usepackage{amsmath}
\usepackage{nccmath}
\usepackage{amsthm}
\usepackage{amssymb}
\usepackage{mathrsfs}
\usepackage{graphicx}
\usepackage{caption}
\usepackage{subcaption}
\usepackage{float} 

\usepackage{empheq}

\newcommand{\be}{\begin{equation}}
\newcommand{\ee}{\end{equation}}
\def\lab{\label}

\begin{document}

\title{\bf Non-Riemannian gravity actions from double field theory }
\author{A.D. Gallegos, U. G\"ursoy, S. Verma and N. Zinnato}
\affiliation{Institute for Theoretical Physics, Utrecht University, Leuvenlaan 4, 3584 CE Utrecht, The Netherlands
\vspace{.5cm}}
\date{\today}

\begin{abstract}
\vspace{.3cm}
Non-Riemannian gravitational theories suggest alternative avenues to understand properties of quantum gravity and provide a concrete setting to study condensed matter systems with non-relativistic symmetry. Derivation of an action principle for these theories generally proved challenging for various reasons. In this technical note, we employ the formulation of double field theory to construct actions for a variety of such theories. This formulation helps removing ambiguities in the corresponding equations of motion. In particular, we embed Torsional Newton-Cartan gravity, Carrollian gravity and String Newton-Cartan gravity in double field theory, derive their actions and compare with the previously obtained results in literature. 
\end{abstract}

\maketitle

{
  \hypersetup{linkcolor=black}
  \tableofcontents
}


\section{Introduction and Summary}\label{sec::intro}

Non-Riemannian gravitational theories have attracted renewed interest  \cite{Kiritsis,Hansen:2020pqs,deBoer,deBoer2} mainly due to the role they might play as an alternative route to understanding properties of quantum gravity \cite{Harmark2014,Hansen:2019vqf,Niels2017}. However, despite various attempts, providing actions for non-relativistic gravity based on local Galilean and Carrollian symmetries proved difficult. It was not until recently that successful attempts  have been made \cite{Bergshoeff:2018vfn,ActionPrinciple, Hansen:2020pqs} for certain classes and limits of non-relativistic geometries. Importance of understanding dynamics of non-Riemannian gravity is underpinned by the far-reaching possibilities this entails: theories based on Galilean symmetry play a role on truncations of string theory  \cite{Gomis:2000bd,Bergshoeff:2019pij,Niels2018}, post-Newtonian physics  \cite{Dautcourt_1997,Dieter,Dieter2,Tichy:2011te}, and provide a natural setting to study response  in condensed matter systems with Galilean symmetry \cite{Son,Hartong2014,MTaylor,Jay}.  Carrollian symmetry, on the other hand, is relevant to description of excitations in the near horizon geometry of black holes \cite{Duval:2014uoa,Damour,Donnay:2019jiz}, and is instrumental in flat space holography \cite{flatHol1,flatHol2}. 

In the particular cases of the Bargmann and Carroll algebras it has been known that the corresponding algebra of generators of the non-Riemannian space-time symmetries can be constructed via null reduction of a parent relativistic theory with a null isometry \cite{Null1,Null2,Duval:1990hj,Duval:1984cj}. In the context of string theory they can be further related by means of T-duality transformations \cite{Bergshoeff:2018yvt,LEE2014134,Morand:2017fnv,Kluson:2018egd}. When T-duality is applied in the null direction of the parent relativistic theory a mapping between non-Riemannian geometries, in particular Torsional Newton-Cartan (TNC) or Carroll type, can be established \cite{Gomis:2000bd,Bergshoeff:2018yvt,Morand:2017fnv}. We note that this can be done for algebras obtained via group contractions of the Poincar\'e group, while algebras obtained from large speed of light expansions, i.e. expansions in $1/c$,  will fall out of the scope of this construction \cite{Hansen:2019vqf,Hansen:2020pqs}. 

This seems to indicate that we should be able to describe both type of geometries, Riemannian and non-Riemannian, in a T-invariant formulation of gravity. Indeed, such a formulation exists. It is based on doubling the degrees of freedom by treating D space-time coordinates and the corresponding D space-time momenta of the compact directions on equal footing  \cite{Hull_2009,Hohm:2010jy,Hohm:2010pp, Siegel:1993xq,Siegel:1993th}. This results in a local $O(D,D)$ invariant theory and the aforementioned T-duality transformation becomes an $O(D,D)$ rotation. This  double field theory (DFT) is based on a generalized metric $\mathcal{H}$ and a generalized dilaton $d$ encompassing the degrees of freedom in the NS sector of string theory, namely the matter content of the theory consists of metric, the Kalb-Ramond field and dilaton. This generalized metric $\mathcal{H}$ is required to be an $O(D,D)$ tensor.
Parametrization of the generalized metric in terms of the original relativistic content $\{g, B,\phi \}$ is known since the postulation of double field theory as a T-duality invariant generalization of string gravity  \cite{Hohm:2010pp}. Importantly however, the generalized metric is not restricted to this form and in particular admits non-Riemannian parametrizations, where the TNC and Carroll limits appear as particular cases  \cite{Morand:2017fnv}. This means that the gravitional dynamics of such non-Riemannian geometries can be obtained by simply considering the double field equations of motion\footnote{These equations can also be unified into a single master equation, as shown in  \cite{Park:2015bza, Angus:2018mep}. We would like to thank Jeong-Hyuck Park for pointing this out.} 
\begin{align}\label{DFTeqs}
	\mathcal{R}_{MN} &= 0 ,&\mathcal{R}&=0
\end{align} 
with $\mathcal{R}\neq \eta^{MN}\mathcal{R}_{MN}$ being the generalization of the Ricci tensor and Ricci scalar to a local $O(D,D)$ geometry. The tensor and curvature scalar appearing in \eqref{DFTeqs} can be written in terms of the generalized metric and dilaton allowing us to fix a non-Riemannian parametrization of $\mathcal{H}$ and obtaining the corresponding gravitational equations of motion. 
 
The goal of this work is to derive the actions and corresponding equations of motion of certain type of non-Riemannian geometries by means of their embedding in double field theory. The relation between DFT and non-Riemannian geometries has already been explored in some detail in the literature, see e.g.   \cite{Ko:2015rha,Berman:2019izh}. We will employ the general parametrization for $\mathcal{H}$ given by Park and Morand  \cite{Morand:2017fnv} and derive the corresponding general equations of motion. We will then specify to particular cases of interest: Torsional Newton-Cartan (TNC) theory, Carrollian theory and the string Newton-Cartan (SNC) theory. Some or part of these equations of motion have been obtained from world-sheet beta-functions of  string theory  \cite{Gomis:2019zyu,Gallegos:2019icg,YanYu} or from reductions of ordinary Einstein's equations. However, in all of these approaches some extra geometric constraints arise. DFT formulation is free of these constraints which will allow us to generalize and complete the existing studies.

The plan of the paper is as follows. In section \ref{DFTsec} we give a brief introduction to double field theory, presenting the basic tensors which will be used to construct the non-relativistic actions. In sections \ref{TNCsec}, \ref{CarSec} and \ref{SNCsec} we determine, respectively, embeddings of TNC, Carroll and SNC theories in double field theory. Using these embeddings we write down the respective actions and compute the equations of motion. In section \ref{CompSec} we compare and discuss the equations of motion found in the current work with the ones already found in the literature. Our work should be viewed as a compendium of useful results on non-Riemannian gravity theories rather than an extensive discussion of their physical properties. 

\subsection*{Notation and conventions}

Throughout the paper greek indices $\left\{\mu,\nu,\rho,\dots\right\}$ denote curved spacetime directions of TNC, Carroll and SNC. The first letters of the latin alphabet (both lowercase and uppercase) refer to flat directions, e.g. the TNC inverse transverse metric is given by $h^{\mu\nu}=e^\mu_a e^\nu_b \delta^{ab}$ and the SNC timelike vielbein is $\tau_\mu^A$. Capital latin letters from the middle of the alphabet $\left\{M,N,\dots\right\}$ are reserved for the DFT directions, e.g. for TNC we will have  $M=0,1,\dots,2d+1,$ where $d=D-1$ is the dimension of the TNC spacetime. Their lowercase versions refer to half of the DFT directions, i.e. $m=0,1,\dots,d $ for the TNC case.

The Riemann tensor is always defined via the action of the commutator of covariant derivatives on an arbitrary vector:
\be
\left[D_\mu, D_\nu\right] A_\rho = -R^{\lambda}_{\ \rho \mu\nu} A_\lambda - 2\Gamma^{\lambda}_{[\mu\nu]} D_\lambda A_\rho
\ee
where $\Gamma^\rho_{\mu\nu}$ is the connection associated to $D_\mu$.

\section{Doubled Gravity} \label{DFTsec}

In this section we review the necessary ingredients of double field gravity. We will not be thorough in our discussion and will only discuss the key ingredients. For a broader discussion see for example  \cite{Hull_2009,Hohm:2010jy, Hohm:2010pp, Morand:2017fnv, Hohm:2011si,ParkRiemann}. A double field theory of gravity should be invariant under $O(D,D)$ rotations and under double diffeomorphisms. As mentioned in the introduction the basic ingredients are the generalized metric $\mathcal{H}$,  generalized dilaton $d$, and the $O(D,D)$ invariant metric $\eta$, given below. Coordinates on double geometry are denoted by $X^M$ and can be decomposed as $X^M= ( X^\mu,\bar X_\nu)$ with the invariant metric
\begin{align}\lab{eta}
	\eta^{MN}= \begin{pmatrix}
		0 & 1 \\
		1 & 0
	\end{pmatrix}\, .
\end{align}
Double diffeomorphisms are generated via the generalized Lie derivative $\mathcal{\hat L}_\xi$ acting on an arbitrary tensor density with weight $\omega$ as \cite{Morand:2017fnv}
\begin{align}\label{generalizedLie}
	\mathcal{L}_\xi T_{M_1 .. M_n} &= \xi^N \partial_N T_{M_1 ... M_n} + \omega \partial_N \xi^N T_{M_1 ... M_n} + \sum_{i=1}^n \left( \partial_{M_i}\xi_N - \partial_N \xi_{M_i} \right) T_{M_1 ... M_{i-1}\hphantom{N}M_{i+1} ... M_n} ^{\hphantom{M_1 ... M_{i-1}}N}
\end{align}
with $\xi^M=(\xi^\mu, \tilde \xi_\mu )$ a generalized vector. The form \eqref{generalizedLie} was originally devised such that it reduces to the one form symmetry for the $B$-field in addition to the standard Lie derivative $\mathcal{ L}_\xi$ on the $X^\mu$ coordinates after the so called ``section'' or strong constraint $\bar \partial^\mu=0$ has been imposed  \cite{DFTReview}. This condition arises from the covariant constraint $\eta_{MN} \partial^M \partial^N = 0$ introduced to reduce the degrees of freedom to their original value and in fact follows from the requirement that the Lie derivatives form a closed symmetry algebra\footnote{We thank Chris Blair for pointing this out to us.}.  Under \eqref{generalizedLie} it can be shown that after exponentiation the generalized dilaton,
 $e^{-2d}$, will act as a scalar density of unit weight and consequently as the integral measure, while, by construction, $\mathcal{H}$ will be an $O(D,D)$ symmetric tensor  satisfying
\begin{align}\label{compD}
	\mathcal{H}_{AC} \eta^{CD} \mathcal{H}_{BD} &= \eta_{AB}.
\end{align}

\subsection{Generalized dilaton and Metric }

When considering a Riemannian manifold the generalized dilaton has a simple expression in terms of the usual dilaton $\phi$ and ordinary ``undoubled'' metric $g_{\mu\nu}$ which will be useful for example when considering TNC and Carrollian geometries: 
\be
e^{-2 d} = e^{-2\phi} \sqrt{-\det g_{\mu\nu} }\equiv e.
\ee

The most general form of $\mathcal{H}$ that is symmetric and compatible with \eqref{compD} is classified by two non-negative integers, $(n,\bar n)$ with $n+\bar n < D$, and is of the form  \cite{Morand:2017fnv} 
\begin{align}\label{parametrization}
	\mathcal{H}_{MN} &= \begin{pmatrix}
	 K_{\mu\nu} - \mathcal{B}_{\mu \rho} H^{\rho \sigma} \mathcal{B}_{\sigma \nu} + 2 x^i_{(\mu} \mathcal{B}_{\nu)\rho} y^\rho_i - 2 \bar x^{\bar{\imath}}_{(\mu} \mathcal{B}_{\nu) \rho}\bar y^\rho_{\bar{\imath}} 	 & & - H^{\nu \rho} \mathcal{B}_{\rho \mu} + y^\nu_i x^i_\mu - \bar y^\nu_{\bar{\imath}} \bar x^{\bar{\imath}}_{\mu} \\
		 - H^{\mu \rho} \mathcal{B}_{\rho \nu} + y^\nu_i x^i_\mu - \bar y^\nu_{\bar{\imath}} \bar x^{\bar{\imath}}_{\mu} & & H^{\mu \nu}
	\end{pmatrix}
\end{align}
with $1\leq i \leq n$ and $1 \leq \bar{\imath} \leq \bar n$. Here, $\mathcal{B}$ is a skew-symmetric matrix that is identified with the Kalb-Ramond field of string theory, while  $K$ and $H$ are symmetric matrices\footnote{These matrices are identified with the target space metric and inverse metric respectively, when the target space is Riemannian}. They have  $n+\bar n$  null eigenvalues each,
\begin{align}\label{kernels}
	 H^{\mu \nu}  x^{ i}_\nu=H^{\mu \nu} \bar x^{\bar{\imath}}_\nu =0 ,\qquad  K_{\mu \nu} y^\nu_j = K_{\mu \nu} \bar y^\nu_{\bar{\jmath}} = 0 \, ,
\end{align}
The corresponding null eigenvectors of $H$ and $K$ are denoted as $x$, $\bar x$ and $y$, $\bar y$ respectively. They are subject to the following completeness relation 
\begin{align}\label{completness}
	H^{\mu \rho} K_{\rho \nu} + y^\mu_i x^i_\nu + \bar y^\mu_{\bar{\imath}} \bar x^{\bar{\imath}}_\nu = \delta^\mu_\nu 
\end{align}
from which the following identities can be inferred 
\begin{align}\label{algId}
	y^\mu_i x^j_\mu &= \delta^j_i , \quad \bar y^\mu_{\bar{\imath}} \bar x^{\bar{\jmath}}_\mu = \delta^{\bar{\jmath}}_{\bar{\imath}} , \quad y^\mu_i \bar x^{\bar{\jmath}}_\mu = \bar y^\mu_{\bar{\imath}} x^j_{\mu} = 0, \quad H^{\rho \mu} K_{\mu \nu} H^{\nu \sigma} = H^{\rho \sigma}, \quad K_{\rho \mu} H^{\mu \nu} K_{\nu \sigma} = K_{\rho \sigma}
\end{align}
%
Once the ``section'' condition is imposed, the generalized Lie derivative \eqref{generalizedLie} reduces to (up to $GL(n)\times GL(\bar n)$ transformations and Milne shifts)

\begin{align}
	\delta x^i_\mu &= \mathcal{L}_\xi x^i_\mu \, , \quad &\delta \bar x^{\bar{\imath}}_{\mu} &= \mathcal{L}_\xi \bar x^{\bar{\imath}}_{\mu} \, , \quad &\delta y^\nu_j &= \mathcal{L}_\xi y^\nu_j \, , \quad &\delta \bar y^\nu_{\bar{\jmath}} &= \mathcal{L}_\xi \bar y^\nu_{\bar{\jmath}} \, , \\ \nonumber
	\delta H^{\mu \nu} &= \mathcal{L}_\xi H^{\mu \nu} \, ,\quad &\delta K_{\mu \nu} &= \mathcal{L}_\xi K_{\mu \nu} \, , \\
	\nonumber \delta \mathcal{B}_{\mu \nu} &= \mathcal{L}_\xi \mathcal{B}_{\mu \nu} + \partial_\mu \tilde \xi_\nu - \partial_\nu \tilde \xi_\mu 
\end{align}
Note that the trace $\mathcal{H}^M_M = 2 (n - \bar n)$\footnote{In  \cite{Park:2020ixf} it was shown that, upon BRST quantization, a critical bosonic string theory can only be anomaly-free if the trace of the generalized metric is zero, i.e. we have to impose $n=\bar n$ at the quantum level for a critical theory to be consistent. } is an $O(D,D)$ invariant scalar and also that the $B$-field acts as an $O(D,D)$ transformation, i.e. its contribution to the generalized metric can be factorized as follows:
\begin{align}\label{conjugation}
	\mathcal{H}_{AB}= \begin{pmatrix}
		1 & \mathcal{B} \\
		0 & 1
	\end{pmatrix} \begin{pmatrix}
		K& Z \\
		Z^T & H  
	\end{pmatrix} \begin{pmatrix}
		1 &0 \\
		 -  \mathcal{B} & 1
	\end{pmatrix}
\end{align}
where we defined
\be
Z^\mu_{\ \nu} \equiv y^\mu_i x_\nu^i - \bar y^\mu_{\bar{\imath}} \bar  x_\nu^{\bar{\imath}}.
\ee
The generalized metric \eqref{parametrization} as well as the relations \eqref{kernels}-\eqref{algId} are invariant under the $GL(n)\times GL(\bar n)$ rotations 

\begin{align}
	\left(x^i_\mu, y^\mu_i , \bar x^{\bar{\imath}}_\mu, \bar y^\nu_{\bar{\imath}} \right) \rightarrow \left( x^j_\mu R^i_j, \left(R^{-1} \right)_i^{\hphantom{i}j} y^\nu_j, \bar x^{\bar{\jmath}}_\mu \bar R_{\bar{\jmath}}^{\hphantom{\bar{\jmath}} \bar{\imath}} , \left(\bar R ^{-1}\right)_{\bar{\imath}}^{\hphantom{\bar{\imath}} \bar{\jmath}} \bar y^\nu_j \right)\, ,
\end{align}
and under the generalized shift symmetry  

\begin{align} \label{shiftDFT}
\begin{split}
\left(y^\mu_i\right)' &= y_i^\mu +V^{\mu}_ i\\
\left(\bar y^\mu_{\bar{\imath}}\right)' &= 	\bar y_{\bar{\imath}}^\mu + \bar V^{\mu}_{\bar{\imath}} \\
\left(K_{\mu\nu}\right)' &= K_{\mu\nu}-2x^i_{(\mu}K_{\nu)\rho}  V^\rho_{i}-2\bar x^{\bar{\imath}}_{(\mu}K_{\nu)\rho}\bar V^\rho _{\bar{\imath}} +\left(x_\mu^i  V_{\rho i} +\bar x^{\bar{\imath}}_\mu \bar V_{\rho \bar{\imath}}  \right) \left(x^i_\nu  V^\rho_i +\bar x^{\bar{\imath}}_\nu\bar V^\rho_{\bar{\imath}} \right)  \\
\left(\mathcal{B}_{\mu\nu}\right)' &= \mathcal{B}_{\mu\nu}-2x^i_{[\mu}  V_{\nu]i}+2\bar x^{\bar{\imath}}_{[\mu}\bar  V_{\nu]\bar{\imath}} + 2x^i_{[\mu}\bar x^{\bar{\jmath}}_{\nu]}\left(y_i^\rho \bar V_{\rho \bar{\jmath}} +\bar y_{\bar{\jmath}}^\rho   V_{\rho i} +V_{\rho i} \bar V^\rho_{\bar{\jmath}} \right)
\end{split}
\end{align}
with $V_{\mu i}$ and $\bar V_{\mu \bar{\imath}}$ being the transformation parameters and we defined $V^\mu_{i}\equiv H^{\mu\rho}V_{\rho i}, \bar V^\mu_{\bar{\imath}}\equiv H^{\mu\rho}\bar V_{\rho \bar{\imath}}$. 

Finally we note that in the present work we do not include a cosmological constant term in the DFT action. However this generalization is straightforward as one only needs to add a term proportional to $e^{-2d}\Lambda_{DFT}$ to the action \cite{Park:2019hbc}. This term could potentially be important when considering non-relativistic holography.

\subsection{Connection and Curvature}

In analogy with general relativity we will introduce a connection $\Gamma_{CAB}$  that will allow us to covariantize derivative interactions. The following unique Christoffel connection is found \cite{ParkRiemann}  
\begin{align}
\begin{split}\label{connection}
		\Gamma_{CAB} &= 2 \left( P \partial_C P \bar P \right)_{[AB]} + 2 \left( \bar P_{[A}^{\hphantom{[A} D} \bar P_{B]}^{\hphantom{B]}E}- P_{[A}^{\hphantom{[A}D} P_{B]}^{\hphantom{B]}E} \right) \partial_D P_{EC} \\
		&\hphantom{=} - 4 \left( \frac{1}{P_M^{\hphantom{M}M}-1} P_{C[A} P_{B]}^{\hphantom{B]}D} + \frac{1}{\bar P_{M}^{\hphantom{M}M}-1} \bar P_{C[A]} \bar P_{B]}^{\hphantom{B]}D}  \right) \left( \partial_D d + \left(P \partial^E P \bar P \right)_{[ED]}\right)
\end{split}
\end{align} 
with $P_{MN}$ and $\bar P_{MN}$ the projector operators defined as 

\begin{align}
	P_{MN}&=\frac{1}{2} \left(\eta_{MN} + \mathcal{H}_{MN} \right)  \, , \\
	\bar P_{MN} &= \frac{1}{2} \left( \eta_{MN} - \mathcal{H}_{MN} \right) \, ,
\end{align}
and satisfying the standard properties $P^2 = P, \bar P^2 = \bar P, P \bar P =0, P + \bar P = 1,$ and with the corresponding traces  

\begin{align}
	P_{M}^{\hphantom{M}M}=D+n - \bar n \, ,\qquad \bar P_M^{\hphantom{M} M} =D-n + \bar n
\end{align}

The connection \eqref{connection} is determined uniquely after imposing compatibility with $\mathcal{H}$, $d$, and the Lie derivative \eqref{generalizedLie} as well as the additional set of projection constraints\footnote{If these projections are not enforced the connection can not be fully determined \cite{DFTReview}. However the relevant covariant curvatures constructed from it will be unique.}
\begin{align}
	\mathcal{P}_{ABC}^{\hphantom{ABC}DEF}\Gamma_{DEF}& =0 ,&  \mathcal{\bar P}_{ABC}^{\hphantom{ABC}DEF}\Gamma_{DEF}& =0 
\end{align}
 with 
 \begin{align}
 	\mathcal{P}_{ABC}^{\hphantom{ABC}DEF} \equiv P_A^{\hphantom{A}D} P_{[B}^{\hphantom{B]}[E}P_{C]}^{\hphantom{C]} F]} + \frac{2}{P_M^{\hphantom{M}M}-1}  P_{A[B}P_{C]}^{\hphantom{C]}[E}P^{F]D} \, , \\
 		\mathcal{\bar P}_{ABC}^{\hphantom{ABC}DEF} \equiv \bar P_A^{\hphantom{A}D} \bar P_{[B}^{\hphantom{B]}[E}\bar P_{C]}^{\hphantom{C]} F]} + \frac{2}{\bar P_M^{\hphantom{M}M}-1}  \bar P_{A[B}\bar P_{C]}^{\hphantom{C]}[E}\bar P^{F]D} \, .
 \end{align}
A field strength $R_{ABCD}$ for the connection $\Gamma$ can be constructed as usual 

\begin{align}\label{fieldStrength}
	R_{CDAB} &= \partial_A \Gamma_{BCD} - \partial_B \Gamma_{ACD} + \Gamma_{AC}^{\hphantom{AC}E}\Gamma_{BED} - \Gamma_{BC}^{\hphantom{BC}E} \Gamma_{AED}
\end{align} 
 However \eqref{fieldStrength} is not a covariant object, in fact no fully covariant four-index Riemann curvature can be constructed in DFT \cite{DFTReview,Hohm:2011si}. Nevertheless we can build the semi-covariant curvature \cite{Hohm:2011si,DFTReview,ParkRiemann} $\mathcal{R}_{ABCD}$
 
 \begin{align}
 	\mathcal{R}_{ABCD} &= \frac{1}{2} \left(R_{ABCD} + R_{CDAB} - \Gamma^E_{\hphantom{E}AB}\Gamma_{ECD} \right)\, , 
 \end{align}
 satisfying symmetry properties
 
 \begin{align}\label{curvatureRie}
 	\mathcal{R}_{ABCD}= \mathcal{R}_{CDAB} = \mathcal{R}_{[AB][CD]}\, , \qquad \mathcal{R}_{A[BCD]}=0
 \end{align}
 and the semi-covariant transformation rule 
 \begin{eqnarray}
 	\delta_\xi \mathcal{R}_{ABCD} &=& \mathcal{\hat  L}_\xi \mathcal{R}_{ABCD} + 2 \nabla_{[A} \left( \left(\mathcal{P} + \mathcal{\bar P} \right)_{B][CD]}^{\hphantom{B][CD]} EFG} \partial_E \partial_F \xi_G  \right) \nonumber  \\ 
	{}&&+ 2 \nabla_{[C} \left( \left(\mathcal{P} 
	+ \mathcal{\bar P} \right)_{D][AB]}^{\hphantom{D][CD]} EFG} \partial_E \partial_F \xi_G  \right) \, .
 \end{eqnarray}
Even though the curvature  defined in \eqref{curvatureRie} is not covariant we can build the following covariant contraction
 \begin{align}\label{curvature1}
 	\mathcal{R} &\equiv \left( P^{AC} P^{BD} - \bar P^{AC} \bar P^{BD} \right) \mathcal{R}_{ABCD}
 \end{align}
which we will call the \textit{doubled Ricci scalar}. It can be shown that  it can be written in terms of the double fields $\mathcal{H}$ and $d$ as 
	\begin{align}
\mathcal{R} &= 4\mathcal{H}^{MN}\partial_M \partial_N d -\partial_M \partial_N \mathcal{H}^{MN}-4\mathcal{H}^{MN}\partial_{M}d\partial_N d+4\partial_M \mathcal{H}^{MN}\partial_N d \nonumber \\
&\quad +\frac{1}{8}\mathcal{H}^{MN}\partial_M \mathcal{H}^{KL}\partial_N \mathcal{H}_{KL}-\frac{1}{2}\mathcal{H}^{MN}\partial_M \mathcal{H}^{KL} \partial_K \mathcal{H}_{NL}
\end{align} 
Equipped with \eqref{curvature1}, we can write down the  DFT equivalent of the Einstein-Hilbert action
\begin{align}\label{action}
	S_{DFT} &= \int d^Dx d^D \tilde x\,  e^{-2d} \mathcal{R} .
\end{align}
The equations of motion are then found by varying \eqref{action}, and given by\footnote{It is important to note that the variation cannot be performed using the parametrization \eqref{parametrization} for fixed $(n,\bar n)$. This would miss  $n \times \bar n$ equations  \cite{Cho:2019ofr}.}
 \begin{align}\label{varAct}
 	\delta S &= \int d^D x d^D \tilde x e^{-2d} \left[ \delta \mathcal{H}^{M N} \mathcal{K}_{M N} - 2 \mathcal{R} \delta d   \right]
 \end{align} 
 with $\mathcal{K}_{M N }$ defined as 
 \begin{align}
\begin{split}
 	\mathcal{K}_{MN} &= \frac{1}{8} \partial_M \mathcal{H}^{KL} \partial_N \mathcal{H}_{KL} - \frac{1}{4} \left( \partial_L - 2  \left(\partial_L d \right) \right) \left(\mathcal{H}^{LK} \partial_K \mathcal{H}_{MN} \right) + 2 \partial_M \partial_N d \\
		&\hphantom{=} - \frac{1}{2} \partial_{(M} \mathcal{H}^{KL} \partial_L \mathcal{H}_{N)K} + \frac{1}{2} \left( \partial_L - 2 \left( \partial_L d \right) \right) \left( \mathcal{H}^{KL} \partial_{(M} \mathcal{H}_{N)K} + \mathcal{H}^{K}_{(M} \partial_K \mathcal{H}^L_{N)} \right)\, . 
\end{split}
 \end{align}
 We need to ensure $\delta H^{MN}$ satisfies \eqref{compD}, this can be achieved as long as we assume the variation takes the form  \cite{Hohm:2010pp} 
 
 \begin{align}\label{varForm}
 	\delta H^{MN} &=P^M_L \delta\mathcal{M}^{LK} \bar P^N_K + \bar P^M_L \delta \mathcal{M}^{LK} P^N_K \, .
 \end{align}
 After assuming the form \eqref{varForm} we find that the equations of motion associated to the action \eqref{action} are nothing but\footnote{The same result could also be obtained by using the following property of the Riemann tensor  \cite{Park:2012xn}:
 $$
 \delta \mathcal{R}_{MNPQ}=\nabla_{[M}\delta \Gamma_{N]PQ}+\nabla_{[P}\delta \Gamma_{Q]MN}.
 $$}
 
 \begin{align}\label{eqm}
 \begin{split}
 	 	\mathcal{R}_{MN} &=0 \, , \\
 	\mathcal{R} &= 0
 \end{split}
 \end{align}
 where the double Ricci  tensor is now expressed in terms of $\mathcal{K}$:
 \begin{align}
 	\mathcal{R}_{MN} &= \bar P^L_M \mathcal{K}_{LK} P^K_N + P^L_M \mathcal{K}_{LK} \bar P^K_N.
 \end{align}
Calling $\mathcal{R}_{MN}$ as generalized Ricci tensor is appropriate as it can be related to the semi-covariant curvature via $\mathcal{R}_{MN}= P^R_M \mathcal{R}^A_{\hphantom{A}RAS} \bar P^S_N$. Note however, $\mathcal{R}\neq \eta^{MN}\mathcal{R}_{MN}$=0. 

In the following sections we will consider action \eqref{action} and equations of motion \eqref{eqm}  for some choices of \eqref{parametrization}, in particular TNC, Carroll and SNC parametrizations. This will allow us to write non-Riemannian gravitational equations of motion by means of the parametrization \eqref{parametrization}. Our usage of the relation between DFT and the previously mentioned non-relativistic geometries will not go further than \eqref{parametrization}. A detailed analysis pertaining the symmetries and related aspects of non-Riemann geometry following from \eqref{parametrization} has been done by Blair et al. in  \cite{newDFT}. 

\section{Type I TNC geometry} \label{TNCsec}
\label{TNC-Geometry}

\subsection{Basics}

A covariant treatment of Galilean symmetry was first obtained by Cartan  \cite{Cartan1923, Cartan2, Friedrichs} which later lead to the Newton-Cartan (NC) geometry as the underlying structure of classical Newtonian gravity. Subsequent work  \cite{Inonu1952, Bargmann:1954gh, LevyLeblond:1967zz, LeblondBook1, Duval_1993} clarified the algebra of  spacetime transformations  and its representation theory that underlies NC geometry. In particular it was shown in  \cite{Andringa:2010it} that the NC geometry follows from gauging the Bargmann algebra, that is, the U(1) central extension of Galilean boost algebra, together with translations and rotations. Generalization of the Newton-Carton geometry to include torsion was obtained in  \cite{Niels2014, connection1}, and this extension was coined ``torsional Newton-Cartan" (TNC) geometry\footnote{In fact, it was argued that it is necessary to include torsion in this theory, see  \cite{connection1}.}. A crucial element in this geometry is the U(1) gauge symmetry corresponding to the aforementioned central charge and physically related to mass conservation. Non-relativistic gravity has recently been studied in the context of non-relativistic effective actions  \cite{Son}, non-relativistic holography  \cite{Hartong2014}, post-Newtonian expansions of general relativity  \cite{Dieter}, and more recently in the context of string theory  \cite{Niels2018,Gallegos:2019icg,Harmark:2019upf}. 
We remark that in this paper we will not consider Type II Newton-Cartan geometries, which arise from a $1/c$ expansion of General Relativity  \cite{Dieter, Hansen:2019vqf, Hansen:2020pqs}.

TNC geometries are characterized by local Galilean symmetry of its tangent space. The non-relativistic spacetime is embedded using a set of spatial frames  $e^a_\mu$ used to define a spatial (or transverse) metric
\begin{align}
	h_{\mu \nu} &= e^a_\mu e^b_\nu \delta_{ab}\, ,
\end{align}
 and a temporal frame $\tau_\mu$ used as a universal one-form clock. The local Galilean symmetry can be extended by including a local $U(1)$ transformation through the  introduction of a mass one-form $m_\mu$. This centrally extended local Galilean symmetry acts on the metric complex $\{e^a_\mu, \tau_\mu  \}$ and the $U(1)$ connection $m_\mu$ as 
\begin{align}
 	\begin{split}\label{transfFrames}
 		\delta e^a_\mu &= \mathcal{L}_\xi e^a_\mu + \lambda^a \tau_\mu + \lambda^a_{\hphantom{b} b} e^b_\mu \,, \\
 		\delta \tau_\mu &= \mathcal{L}_\xi \tau_\mu \, , \\
 		\delta m_\mu &= \mathcal{L}_\xi m_\mu + \lambda_a e^a_\mu + \partial_\mu \sigma \, ,
 	\end{split}
 \end{align} 
with $\lambda^a$ a local Galilean boost parameter, $\lambda^{ab}$ a local parameter of rotations, $\sigma$ a $U(1)$ gauge transformation parameter, and $\xi^\mu$ parametrizing diffeomorphisms. The square matrix $(\tau_\mu,e^a_\mu)$ has an inverse $(-\upsilon^\mu, e^\mu_a)$ where these inverse frames are orthogonal
\be\lab{orthTNC}
		\upsilon^\mu \tau_\mu = -1 \, , \qquad
		e^\mu_a \tau_\mu = 0 \, , \qquad
		e^a_\mu \upsilon^\mu = 0 \, ,
\ee
and complete
 \be\label{completFrames}
  e^a_\nu e_a^\mu - \upsilon^\mu \tau_\nu =\delta ^\mu_\nu	\, .
 \ee
Making use of inverse spatial frames $e^\mu_a$, one can define an inverse spatial metric as 
\begin{align}
	h^{\mu \nu} &= e^\mu_a e^\nu_b \delta^{ab} \, ,
\end{align}
such that the completeness relation can be rewritten as
\be
  h^{\mu\rho}h_{\rho\nu}- \upsilon^\mu \tau_\nu =\delta ^\mu_\nu\, .	
\ee
It is well known that a geometry described by these fields and symmetries can be obtained from \textit{null} reduction of a relativistic theory described by the line element
\be \label{NullTNC}
ds^2= 2\tau_\mu dx^\mu (du -m_\nu dx^\nu)+h_{\mu\nu}dx^\mu dx^\nu,
\ee
where $u$ is the direction associated to the null isometry, i.e. $\partial_u$ is a Killing vector. Note that the null reduction of the relativistic connection will not give exactly the connection we use in this text,  \eqref{TNCconn}, however the difference between the two is simply given by the torsion component.

Instead of the set of variables $\{\tau_\mu, \upsilon^\nu,h^{\mu \nu}, h_{\mu \nu}, m_\mu \}$ it will be more convenient for us to work with the boost-invariant set $\{\tau_\mu,\hat \upsilon^\mu, \bar h_{\mu \nu}, h^{\mu \nu}, \Phi \}$ where $\{\hat \upsilon^\mu, \bar h_{\mu \nu}, \Phi\}$ are defined as 
\begin{align}
	\begin{split}\lab{hatups}
		\bar h_{\mu \nu} &\equiv h_{\mu \nu} - \tau_\mu m_\nu - \tau_\nu m_\mu,\\
		\hat \upsilon^\mu &\equiv \upsilon^\mu - h^{\mu \nu} m_\nu \\
		\Phi &\equiv - \upsilon^\rho m_\rho +\frac{1}{2}h^{\rho\sigma}m_\rho m_\sigma
	\end{split}
\end{align}
which are the boost-invariant combinations with the physical interpretation of a spatial metric, an ``inverse clock'', and Newton gravitational potential (i.e. the scalar which will appear in Poisson's equation). While these variables are explicitly invariant under Galilean boosts, they still transform under the $U(1)$ extension as 
\be
		\delta_\sigma \bar h_{\mu \nu} = - 2 \tau_{(\mu}\partial_{\nu)}\sigma \, ,  \qquad
		\delta_\sigma \hat \upsilon^\mu = - h^{\mu \nu} \partial_\nu \sigma \, , \qquad
		\delta_\sigma \Phi =- \hat \upsilon^\rho \partial_\rho \sigma \, ,
	\ee
whereas $\{\tau_\mu, h^{\mu \nu} \}$ remain invariant under all local gauge transformations. The Galilean invariant fields are subject to the identities
\be\lab{Galids}
		\bar h_{\mu \rho} h^{\rho \nu} - \hat \upsilon^\nu \tau_\mu = \delta^\nu_\mu \, , \qquad
		\hat \upsilon^\mu \tau_\mu = - 1 \, , \qquad
		\hat \upsilon^\mu \bar h_{\mu \nu} = 2 \Phi \tau_\nu \, .
\ee
The connection we will use is\footnote{See also  \cite{Bernal:2002ph, Bekaert:2014bwa} for a classification of TNC connections.}
\be \label{TNCconn}
\Gamma^\rho_{\mu\nu} = -\hat \upsilon^\rho\partial_\mu\tau_\nu +\frac{1}{2}h^{\rho\sigma}\left(\partial_\mu \bar h_{\nu\sigma}+\partial_\nu \bar h_{\mu\sigma}-\partial_\sigma \bar h_{\mu\nu}\right)
\ee
which is manifestly boost invariant, but not $U(1)$ invariant. Note that this connection is compatible with $h^{\mu\nu}$ and $\tau_\mu$. The antisymmetric part of this connection is poportional to the torsion tensor of TNC:
\be
F_{\mu\nu}\equiv\partial_{[\mu}\tau_{\nu]},
\ee
in terms of which we define the acceleration
\be
a_\mu \equiv \hat \upsilon^{\rho}F_{\rho\mu}.
\ee
In the following section we will reformulate this theory by embedding it in DFT. 
\subsection{Embedding in DFT}

By examining \eqref{NullTNC} we can identify the following metric and inverse relativistic metrics:
\begin{align} \label{RelMetTNC}
	g_{\mu \nu} &= \begin{pmatrix}
		\bar h_{\mu \nu}  &\tau_\mu \\ 
		\tau_\mu & 0
	\end{pmatrix} \, , \qquad &g^{\mu \nu} &= \begin{pmatrix}
		h^{\mu \nu} & -\hat \upsilon^\mu \\
		- \hat \upsilon^\mu & 2 \Phi 
	\end{pmatrix},
\end{align}
which can be embedded in DFT as
\begin{align} \label{GMTNC}
	\mathcal{H}_{MN} &= \begin{pmatrix}
		\bar h_{\mu \nu} & \tau_\mu & 0 & 0 \\
		\tau_\mu &0 & 0 & 0 \\ 
		0 & 0 & h^{\mu \nu} & - \hat \upsilon^\mu \\
		0 & 0 & -\hat \upsilon^\mu & 2 \Phi 
	\end{pmatrix}\, .
\end{align}
We now apply a T-duality transformation to swap the null direction $u$ with the dual null direction $\bar u$,
\begin{align}
	\mathcal{T}^M_{\hphantom{M}N} &= \begin{pmatrix}
		\delta^\mu_{\hphantom{\mu}\nu} & 0 & 0 & 0 \\
		0& 0 & 0 & 1 \\
		0& 0 & \delta_{\mu}^{\hphantom{\mu}\nu } &0 \\
		0 & 1 &0 & 0
	\end{pmatrix},
\end{align}
which results in the following \textit{TNC generalized metric}:
\begin{align} \label{GMTNC2}
	\mathcal{H}_{MN} &= \begin{pmatrix}
		\bar h_{\mu \nu} & 0 & 0 & \tau_\mu \\
		0 & 2 \Phi & - \hat \upsilon^\nu & 0 \\
		0 & - \hat \upsilon^\mu & h^{\mu \nu} & 0 \\
		\tau_\nu & 0 & 0 & 0 
	\end{pmatrix}.
\end{align}
The generalized metrics \eqref{GMTNC} and \eqref{GMTNC2} will produce the same actions and equations of motion, however note that the lower right block in \eqref{GMTNC2} is now degenerate meaning that the latter has a clearer non-relativistic interpretation. As shown in  \cite{Blair:2019qwi}, after adding matter to this parametrization we find that the embedding is given by\footnote{Note that now we are briefly using $m,n$ as indices of the DFT tensors, as a way to avoid introducing new symbols. We have $m=0,1,\dots,d+1$ and $\mu=0,1,\dots,d$ where $d$ is the dimension of the TNC spacetime and $d+1=D$ is the dimension of its uplift, which includes the null direction $u$.  }
\begin{align}\label{TNCDFT1}
K_{mn} &=   \begin{pmatrix}
			h_{\mu\nu} && 0 \\ 
			0 && 0 
		\end{pmatrix}, & 
H^{mn} &=  \begin{pmatrix}
			h^{\mu\nu} && h^{\mu\rho}\aleph_\rho \\ 
			h^{\nu\rho}\aleph_\rho && h^{\rho\sigma}\aleph_\rho\aleph_\sigma 
		\end{pmatrix},&
\mathcal{B}_{mn} &=  \begin{pmatrix}
			\bar B_{\mu\nu} && -m_\mu \\ 
			m_\nu && 0
		\end{pmatrix}
\end{align}
and
\begin{align} 
x_m&=\frac{1}{\sqrt{2}}  \begin{pmatrix}
			\tau_\mu - \aleph_\mu  \\ 
			 1 
		\end{pmatrix},&
\bar x_m& =\frac{1}{\sqrt{2}}  \begin{pmatrix}
			\tau_\mu+ \aleph_\mu  \\ 
			 -1 
		\end{pmatrix},\nonumber\\
y^m& =\frac{1}{\sqrt{2}}  \begin{pmatrix}
			-\upsilon^\mu  \\ 
			 1 -\upsilon^\mu \aleph_\mu
		\end{pmatrix},&
\bar y^m &=\frac{1}{\sqrt{2}}  \begin{pmatrix}
			-\upsilon^\mu  \\ 
			 -1-\upsilon^\mu \aleph_\mu,
		\end{pmatrix}. \label{TNCDFT2}
\end{align}
where $\aleph_\mu \equiv -B_{\mu u}$ appears from the dimensional reduction of the $B$-field. We also define
\be
b_{\mu\nu} \equiv \partial_{[\mu}\aleph_{\nu]}, \qquad \qquad \mathfrak{e}_\mu \equiv \hat 
\upsilon^\rho b_{\rho\mu}.
\ee

The tensors $K_{\mu\nu}$ and $H^{\mu\nu}$ both have $2=1+1$ null eigenvectors and the trace of the generalized metric is $\mathcal{H}^{M}_M=0$, implying that a TNC geometry corresponds to a (1,1) theory in the DFT framework. 

\subsection{Action and equations of motion}
\label{Action-Equations}
 
Having obtained the embedding of TNC geometry in double field theory, we can now immediately write down its action using the DFT action in equation (\ref{action}). 
We first introduce the following notation. Given an arbitrary tensor $A_{\mu\nu\dots}$ we will define for brevity
\begin{equation} \label{UpInd}
A^{\mu\nu\dots}\equiv A_{\rho\sigma\dots} h^{\mu\rho}h^{\nu\sigma}\dots
\end{equation} 
Note that the only TNC fields which naturally have upper indices are $\hat \upsilon^\mu $ and $h^{\mu\nu}$, hence any other tensor with upper indices is to be understood as defined via \eqref{UpInd}.

Now, the TNC action is given by\footnote{It is worth noting that this same action can also be  obtained as the dimensional reduction of the standard NS-NS sector of supergravity 
$$
S_{NS-NS}=\int d^{D }x\, e^{-2\phi}\sqrt{-\det g}\left(\mathcal{R}-\frac{1}{12} H^2+ 4 \left(\partial \phi\right)^2 \right)
$$ 
where $g$ is the Riemannian metric appearing in \eqref{RelMetTNC}, $H_{\mu\nu\rho}$ is the field-strength of the B-field and $D=d+1$ with $d$ the dimension of the TNC spacetime. Note however that this null reduction is not fully consistent, as one needs to impose Poisson's equation on-shell, rather than deriving it from the (null-reduced) action.}
\begin{align}\label{TNCaction}
	\begin{split}
		S =& \int d^d x\,e\, \left[ \mathcal{R}+ \frac{1}{2}a^\mu a_\mu + \frac{1}{2} \mathfrak{e}^\mu \mathfrak{e}_\mu  -4a^\mu D_\mu \phi+4D^\mu\phi D_\mu \phi   - \frac{1}{12} H^{\mu \nu\rho}H_{\mu\nu\rho}\right.\\
		&\hphantom{(((\int d^d x e\,  }   \left.   - \frac{1}{2} \hat \upsilon^\rho H_{\rho\mu\nu } b^{\mu\nu} - \frac{1}{2} \left(F^{\mu \nu} F_{\mu \nu}+b^{\mu\nu}b_{\mu \nu} \right)\Phi  \right]
	\end{split}
\end{align}
with $H=dB$, $b=d\aleph$ and
\begin{equation} \label{TNCmeasure}
e \equiv \sqrt{\frac{\text{det} \bar h}{2 \Phi}} e^{-2\phi}.
\end{equation}
We choose the independent fields of our theory to be $\Phi, \hat \upsilon^\mu, h^{\mu\nu}, B_{\mu\nu}$ and $\aleph_\mu$, see Appendix \ref{AppTNC} for useful identites.

%

The variation of the action \eqref{TNCaction} with respect to $\Phi$ imposes a generalized twistlessness contraint on torsion:
\begin{equation} \label{twistTNC}
F_{\mu\nu}F^{\mu\nu}=b_{\mu\nu}b^{\mu\nu}.
\end{equation}
The equations for the matter fields are
\begin{align} \label{MatterTNC}
D_\mu D^\mu \phi +a^\mu D_\mu\phi -2 D_\mu\phi  D^\mu \phi& =\frac{1}{2}\mathfrak{e}^2 -\frac{1}{12}H_{\mu\nu\rho}H^{\mu\nu\rho} -\frac{1}{2}\hat \upsilon^\lambda H_{\lambda\mu\nu}b^{\mu\nu}-\frac{1}{2}F_{\mu\nu}F^{\mu\nu} \Phi   \\
D^\rho b_{\rho\mu}-2 D^\rho \phi\, b_{\rho\mu}&=\frac{1}{2}F^{\rho\sigma}H_{ \rho \sigma\mu} \\
 D^\rho H_{\rho\mu\nu}+a^\rho H_{\rho\mu\nu}-2D^\rho\phi \,H_{\rho\mu\nu}&=2D_{[\mu}\mathfrak{e}_{\nu]}-2\hat \upsilon^\lambda D_\lambda b_{\mu\nu}+2a_{[\mu}\mathfrak{e}_{\nu]}+4b^\rho_{\hphantom{\rho} [\mu}F_{\nu]\rho}\Phi\nonumber\\
& \quad +\left(2\hat \upsilon^\rho D_\rho \phi- D_\rho \hat \upsilon^\rho \right)b_{\mu\nu},
\end{align}
while the equation for $\tau_\mu$ is
\begin{align} \label{FTNC}
D^\rho F_{\rho\mu}+a^\rho F_{\rho\mu}-2 D^\rho\phi F_{\rho\mu}&=\frac{1}{2}b^{\rho\sigma}H_{ \rho \sigma\mu}+\mathfrak{e}^\rho b_{\rho \mu} . 
\end{align}
The equation obtained from the variation with respect to $h^{\mu\nu}$ requires  more care. First of all, notice that we must have $\tau_\mu \tau_\nu \delta h^{\mu\nu}=0$. This means that the most general variation is given by 
\begin{equation}
\delta h^{\mu\nu} = \left(\Delta_S\right)_\rho^\mu \delta \mathcal{M}^{\rho\nu}+\delta \mathcal{M}^{\mu\rho} \left(\Delta_S\right)_\rho^\nu 
\end{equation}
where $\delta \mathcal{M}^{\mu\nu}$ is an arbitrary symmetric tensor and $\left(\Delta_S\right)_\nu^\mu = h^{\mu\rho}\bar h_{\rho\nu}$.  It also follows that the time-time projection of the equation obtained from this variation will be trivially zero, i.e. we will not be able to obtain Newton's law from this variation.  This will also be true for variation of the Carrollian and SNC  actions and is, in fact, a general property inherited from double field theory  \cite{Cho:2019ofr}. By imposing an ansatz on the generalized metric  and then computing the variation of the resulting action we will end up with $n\times \bar n$ equations less than we would initially expect. However, this problem can be easily avoided by taking the variation of the DFT action \textit{first} and imposing the TNC ansatz on the resulting equations of motion \eqref{DFTeqs}. With this in mind, we find that the variation of \eqref{TNCaction} with respect to $h^{\mu\nu}$ produces the equation
\begin{align} \label{Ne0}
\begin{split}
 \mathcal{R}_{(\mu\nu)} +2 D_{(\mu}D_{\nu)}\phi-\frac{1}{4}h^{\rho\sigma}h^{\lambda\kappa}H_{\mu\rho\lambda}H_{\nu\sigma\kappa}&=\frac{a_\mu a_\nu-\mathfrak{e}_\mu\mathfrak{e}_\nu}{2}-\hat \upsilon^\rho D_{(\mu}F_{\nu)\rho} +\hat \upsilon^{\lambda}h^{\rho\sigma}b_{\rho(\mu}H_{\nu)\lambda\sigma} \\
&\quad -\left(F_{\mu\rho}F_{\nu\sigma}h^{\rho\sigma} -b_{\mu\rho}b_{\nu\sigma}h^{\rho\sigma}\right)\Phi  \\
&\quad+\mathfrak{R}_{\rho\sigma}\left(\Delta_T\right)^\rho_\mu\left(\Delta_T\right)^\sigma_\nu
\end{split}
\end{align}
where we defined
\begin{equation}
\mathfrak{R}_{\mu\nu}\equiv \mathcal{R}_{\mu\nu}+2 D_\mu D_\nu \phi-\frac{1}{4} h^{\rho\sigma}h^{\lambda\kappa}H_{\mu\rho\lambda}H_{\nu\sigma\kappa}+  a^2 \tau_\mu \tau_\nu \Phi -\mathfrak{e}^2 \tau_\mu \tau_\nu \Phi.
\end{equation}
Note that because of the presence of $\mathfrak{R}_{\mu\nu}$ in \eqref{Ne0} the time-time projection of Einstein's equations is identically zero. As we already explained, Newton's law can  be found by imposing the TNC ansatz on the DFT equations \eqref{DFTeqs}. The resulting equation is 
\begin{equation}\label{newt0}
D^\mu D_\mu \Phi +3 a^\mu D_\mu \Phi  + m^2_{\Phi} \Phi -2F_{\mu\nu}F^{\mu\nu}\Phi^2 = \rho_{\mathcal{K}}+\rho_{m}
\end{equation}
with
\begin{align}
m^2_{\Phi} &= a^2+\mathfrak{e}^2+4 a^\mu D_\mu\phi -\hat \upsilon^\rho H_{\rho\mu\nu}b^{\mu\nu},\\
\rho_{\mathcal{K}} &= \hat \upsilon^\mu D_\mu D_\nu \hat \upsilon^\nu +D_\mu \hat \upsilon^\nu D_\nu \hat \upsilon^\mu =-\hat \upsilon^\mu D_\mu \mathcal{K}^\nu_\nu+ \mathcal{K}_{\mu\nu}\mathcal{K}^{\mu\nu}, \\
\rho_m &= \frac{1}{4} \hat \upsilon^\mu \hat \upsilon^\nu H_{\mu\rho\sigma}H_\nu^{\hphantom{\nu} \rho\sigma} -2 \hat \upsilon^\mu \hat\upsilon^\nu D_\mu D_\nu \phi.
\end{align}
Notice that by using the identities
\begin{align}
\hat \upsilon^\mu \hat \upsilon^\nu \mathcal{R}_{\mu\nu} & = -\hat\upsilon^\mu D_\mu D_\nu \hat \upsilon^\nu +\hat \upsilon^\mu D_\nu D_\mu \hat \upsilon^\nu +a^\mu D_\mu \Phi +2 a^2 \Phi, \\
D_\mu \hat \upsilon^\rho D_\rho \hat \upsilon^\nu +\hat \upsilon^\rho D_\mu D_\rho \hat \upsilon^\nu &= h^{\nu \sigma} \left( D_\mu D_\sigma \Phi +2 D_\mu a_\sigma \, \Phi +2a_\sigma D_\mu \Phi   \right), 
\end{align}
we can rewrite  \eqref{newt0} as
\begin{equation}
\hat \upsilon^\mu \hat \upsilon^\nu \mathcal{R}_{\mu\nu}=\frac{1}{4} \hat \upsilon^\mu \hat \upsilon^\nu H_{\mu\rho\sigma}H_\nu^{\hphantom{\nu} \rho\sigma} -2 \hat \upsilon^\mu \hat\upsilon^\nu D_\mu D_\nu \phi -a^2\Phi+\mathfrak{e}^2\Phi
\end{equation}
so that the full  Einstein's equations can compactly be written as
\begin{align}  \label{EinsteinTNC}
\begin{split}
 \mathcal{R}_{(\mu\nu)} +2 D_{(\mu}D_{\nu)}\phi-\frac{1}{4}h^{\rho\sigma}h^{\lambda\kappa}H_{\mu\rho\lambda}H_{\nu\sigma\kappa}&=\frac{a_\mu a_\nu-\mathfrak{e}_\mu\mathfrak{e}_\nu}{2}-\hat \upsilon^\rho D_{(\mu}F_{\nu)\rho} +\hat \upsilon^{\lambda}h^{\rho\sigma}b_{\rho(\mu}H_{\nu)\lambda\sigma} \\
&\quad -\left(F_{\mu\rho}F_{\nu\sigma}h^{\rho\sigma} -b_{\mu\rho}b_{\nu\sigma}h^{\rho\sigma}\right)\Phi.
\end{split}
\end{align}
In summary, the TNC equations of motion are given by 
\begin{align} \label{fullTNC}
F_{\mu\nu}F^{\mu\nu}&=b_{\mu\nu}b^{\mu\nu},\\
D^\rho F_{\rho\mu}+a^\rho F_{\rho\mu}-2 D^\rho\phi F_{\rho\mu}&=\frac{1}{2}b^{\rho\sigma}H_{ \rho \sigma\mu}+\mathfrak{e}^\rho b_{\rho \mu}\\
D^\rho b_{\rho\mu}-2 D^\rho \phi\, b_{\rho\mu}&=\frac{1}{2}F^{\rho\sigma}H_{ \rho \sigma\mu} ,\\
D_\mu D^\mu \phi +a^\mu D_\mu\phi -2 D_\mu\phi  D^\mu \phi& =\frac{1}{2}\mathfrak{e}^2 -\frac{1}{12}H_{\mu\nu\rho}H^{\mu\nu\rho} -\frac{1}{2}\hat \upsilon^\lambda H_{\lambda\mu\nu}b^{\mu\nu}-\frac{1}{2}F_{\mu\nu}F^{\mu\nu} \Phi ,  \\
 D^\rho H_{\rho\mu\nu}+a^\rho H_{\rho\mu\nu}-2D^\rho\phi \,H_{\rho\mu\nu}&=2D_{[\mu}\mathfrak{e}_{\nu]}-2\hat \upsilon^\lambda D_\lambda b_{\mu\nu}+2a_{[\mu}\mathfrak{e}_{\nu]}+4b^\rho_{\hphantom{\rho} [\mu}F_{\nu]\rho}\Phi\nonumber\\
& \quad +\left(2\hat \upsilon^\rho D_\rho \phi- D_\rho \hat \upsilon^\rho \right)b_{\mu\nu},\\
 \mathcal{R}_{(\mu\nu)} +2 D_{(\mu}D_{\nu)}\phi-\frac{1}{4}h^{\rho\sigma}h^{\lambda\kappa}H_{\mu\rho\lambda}H_{\nu\sigma\kappa}&=\frac{a_\mu a_\nu-\mathfrak{e}_\mu\mathfrak{e}_\nu}{2}-\hat \upsilon^\rho D_{(\mu}F_{\nu)\rho} +\hat \upsilon^{\lambda}h^{\rho\sigma}b_{\rho(\mu}H_{\nu)\lambda\sigma}\nonumber \\
&\quad -\left(F_{\mu\rho}F_{\nu\sigma}h^{\rho\sigma} -b_{\mu\rho}b_{\nu\sigma}h^{\rho\sigma}\right)\Phi,
\end{align}
where we remind the reader of the short-hand notation \eqref{UpInd}, as well as the following definitions
\begin{align}
\begin{split}
a_{\mu} &= \hat\upsilon^\rho F_{\rho \mu} = 2\hat \upsilon^\rho\partial_{[\rho}\tau_{\mu]},\\ 
\mathfrak{e}_{\mu} &= \hat\upsilon^\rho b_{\rho \mu} = 2\hat \upsilon^\rho\partial_{[\rho}\aleph_{\mu]}
\end{split}
\end{align}
and $\Phi$ is Newton's potential, not to be confused with the dilaton $\phi$. 

Note that these equations are \textit{manifestly} invariant under almost all transformations described by the Bargmann algebra. The only nontrivial transformation corresponds to the $U(1)$ generator $m_\mu$. However a straightforward (but tedious) computation shows that these equations are indeed invariant under mass $U(1)$, although not manifestly so. We will discuss some properties of these equations and their relation to known results in section \ref{CompSec}.

\section{Carrollian geometry } \label{CarSec}

\subsection{Basics}

The Carroll algebra can be obtained by considering a particular contraction ($c\rightarrow 0$) of the Poincar\'e algebra  \cite{Levy1965,Hartong:2015xda,Bergshoeff:2017btm}. In  \cite{Duval:2014uva} it was suggested that this algebra could play an important role in flat space holography, hence it would be interesting to study how field theories couple to Carrollian spacetime, see e.g.  \cite{Hofman:2014loa, Bagchi:2019clu, Bagchi:2019xfx}. To this end, one needs a gravitational action coupled to matter, and this is precisely what we will compute below.

We will describe Carrollian geometry using the same set of symbols we already introduced for the TNC case above. However, these fields will now transform under  the Carrollian boosts rather than the Galilean boosts:
\begin{equation}
\delta_\mathcal{C} \tau_\mu =\lambda_\mu , \qquad \qquad \delta_\mathcal{C} h^{\mu\nu}= 2h^{\rho (\mu}\upsilon^{\nu)}\lambda_\rho.
\end{equation}
Moreover we replace the gauge field $m_\mu$ by a contravariant vector $M^\mu$, which transforms as
\begin{equation}
\delta M^\mu = \mathcal{L}_\xi M^\mu + e^\mu_a \lambda^a =  \mathcal{L}_\xi M^\mu + h^{\mu\nu} \lambda_\nu. 
\end{equation}
This allows us to build the following manifestly boost-invariant tensors
\begin{align}\label{CarBasics}
\hat \tau_\mu &= \tau_\mu -h_{\mu\nu}M^\nu,\nonumber\\
\hat h^{\mu\nu} &= h^{\mu\nu}-2M^{(\mu}\upsilon^{\nu)} +2\Phi \upsilon^\mu \upsilon^\nu \equiv \bar h^{\mu\nu}+2\Phi \upsilon^\mu \upsilon^\nu,\\
\Phi& = -M^\mu \tau_\mu +\frac{1}{2}h_{\mu\nu}M^\mu M^\nu= -M^\mu \hat\tau_\mu -\frac{1}{2}h_{\mu\nu}M^\mu M^\nu\, ,\nonumber
\end{align}
which have the interpretation as the boost-invariant clock one-form, inverse spatial metric and the Newton potential respectively. They satisfy the following orthogonality and completeness relations 
\begin{equation}
\hat h^{\mu\rho}h_{\rho\nu}-\upsilon^\mu \hat \tau_\nu=\delta^\mu_\nu,\qquad \qquad \hat \tau_\mu \hat h^{\mu\rho}=0=\upsilon^\mu h_{\mu\rho}.
\end{equation}
It is possible to embed a Carrollian geometry in a Lorentzian one just as we did for TNC in \eqref{NullTNC}:
\be\label{NullCar}
ds^2=du(2\Phi\, du -2\hat \tau_\mu dx^\mu) +h_{\mu\nu}dx^\mu dx^\nu
\ee
A connection compatible with $\hat \tau_\mu, \upsilon^\mu, h_{\mu\nu}$ and $\hat h^{\mu\nu}$ can be constructed   \cite{Hartong:2015xda, Bekaert_2018}:
\begin{equation}
\widetilde \Gamma^\rho_{\mu\nu}=-\upsilon^\rho \partial_\mu \hat \tau_\nu +\frac{1}{2}\hat  h^{\mu\lambda}\left(\partial_\mu h_{\nu\lambda}+\partial_\nu h_{\mu\lambda}-\partial_\lambda h_{\mu\nu}\right) - \hat h^{\mu\lambda} \mathcal{K}_{\lambda\mu}\hat\tau_\nu,
\end{equation}
where we introduced the extrinsic curvature
\begin{equation} \label{CarExt}
\mathcal{K}_{\mu\nu}=-\frac{1}{2}\mathcal{L} _{\upsilon} h_{\mu\nu}=-\frac{1}{2}\left( \upsilon^\rho \partial_\rho h_{\mu\nu}+\left(\partial_\mu \upsilon^\rho\right)h_{\rho\nu}+\left(\partial_\nu \upsilon^\rho\right)h_{\mu\rho}\right).
\end{equation}
However, we find it more convenient to use a slightly different connection, 
\begin{equation}
\Gamma^\rho_{\mu\nu}\equiv  \widetilde \Gamma^\rho_{\mu\nu}+\hat h^{\mu\lambda} \mathcal{K}_{\lambda\mu}\hat\tau_\nu.
\end{equation}
Using $\hat \tau_{\mu}\hat h^{\mu\nu}=0$, it is easy to show that this connection is still compatible with $\hat \tau_\mu$ and $\hat h^{\mu\nu}$, but now we have
\begin{equation}
D_\mu \upsilon^\nu= -\hat h^{\nu\lambda}\mathcal{K}_{\lambda\mu},\qquad \qquad D_\rho h_{\mu\nu}=-2\mathcal{K}_{\rho(\mu}\hat \tau_{\nu)}
\end{equation}
where we used  
\begin{equation}
\upsilon^\mu \mathcal{K}_{\mu\nu}=\upsilon^\rho D_\rho \upsilon^\mu=0.
\end{equation}
We also define the following tensors in analogy with TNC:
\be
F_{\mu\nu}\equiv \partial_{[\mu}\hat\tau_{\nu]},\qquad \qquad a_\mu \equiv \upsilon^\rho F_{\rho\mu}.
\ee
\subsection{Embedding in DFT}
Given the relativistic geometry \eqref{NullCar} we can construct the generalized metric
\begin{align} \label{GMCar}
	\mathcal{H}_{MN} &= \begin{pmatrix}
		 h_{\mu \nu} & - \hat \tau_\mu & 0 & 0 \\
		-\hat \tau_\mu & 2  \Phi & 0 & 0 \\ 
		0 & 0 & \bar h^{\mu \nu} & \upsilon^\mu \\
		0 & 0 & \upsilon^\mu & 0
	\end{pmatrix}.
\end{align}
There are clearly some similarities between this metric and the TNC one \eqref{GMTNC}. In fact when the scalars $\Phi^{(\mathcal{C})}$ and $\Phi^{(TNC)}$ are both zero (which implies $M^\mu=m_\mu=0$) it is easy to see that the two generalized metrics become identical up to some obvious identifications. When the two scalars are not zero we still have a relation between the two geometries, but it is a bit more involved.  To see this we can start from \eqref{GMCar} and then apply a T-duality transformation that swaps the Carrollian directions $\mu$ with their dual ones $\bar \mu$ and arrive at  
\begin{align} \label{GMCar2}
	\mathcal{H}_{MN} &\rightarrow  \begin{pmatrix}
		\bar h^{\mu \nu} & 0 & 0 & \upsilon^\mu \\
		0 & 2 \Phi & - \hat\tau_\nu & 0 \\
		0 & - \hat \tau_\mu & h_{\mu \nu} & 0 \\
		\upsilon^\nu & 0 & 0 & 0 
	\end{pmatrix}.
\end{align}
This is equivalent to \eqref{GMTNC2} if we make the identifications
\begin{align} \label{TNCarrDual1}
 \bar h^{\mu\nu}_{(\mathcal{C})}& \leftrightarrow \bar h_{\mu\nu}^{(TNC)}, & \hat \tau_\mu ^{(\mathcal{C})}& \leftrightarrow \hat \upsilon^\mu_{(TNC)}, & \Phi^{(\mathcal{C})}& \leftrightarrow \Phi^{(TNC)},\\
 h_{\mu\nu}^{(\mathcal{C})}& \leftrightarrow h^{\mu\nu}_{(TNC)}, & \upsilon^\mu_{(\mathcal{C})} &\leftrightarrow \tau_\mu^{(TNC)}, & M^\mu_{(\mathcal{C})} &\leftrightarrow m_\mu^{(TNC)}\label{TNCarrDual2}.
\end{align}
This is the same duality that was proposed  and discussed in  \cite{Hartong:2015xda,Duval_2014, Bekaert_2018}, however when working in the DFT framework it is  clear what the interpretation of this duality is, i.e. it is simply a T-duality following from the fact that this theory is embedded in string theory! Note that this duality is mapping two theories that are in principle really different, since one is a non-relativistic theory while the other is an ultra-relativistic theory. In some sense it seems like this T-duality is acting on these spacetimes as $c \leftrightarrow 1/c$ with $c$ being the speed of light. It is also interesting to note that when acting with this transformation on the Carrollian side we are effectively generating massive particles, which will correspond to the eigenstates of the $U(1)$ generator on the TNC side.

Moreover we remark that this duality only maps the TNC generalized metric to the Carrollian one (and vice versa). To map the full actions to one another we would need to transform the partial derivatives as well, i.e. we would need a transformation of the form
\be
(\partial_\mu)^{(Car)}\leftrightarrow (\partial^\mu)^{(TNC)},
\ee
but it is not clear what a partial derivative with upper index means in a non-relativistic (or ultra-relativistic) theory. It will be interesting to explore this duality in more detail in future works, to understand if the two actions can indeed be mapped to each other.

\subsection{Action and equations of motion}
Given an arbitrary tensor $A_{\mu\nu\dots}$ we define
\begin{equation} \label{UpIndC}
A^{\mu\nu\dots}\equiv A_{\rho\sigma\dots} \hat h^{\mu\rho}\hat h^{\nu\sigma}\dots
\end{equation} 
The only Carroll fields which naturally have upper indices are $  \upsilon^\mu, \bar h^{\mu\nu}, \hat h^{\mu\nu}$ and $M^\mu$, hence any other tensor with upper indices is to be understood as defined via \eqref{UpIndC}.

The action for a Carrollian gravitational theory is given by\footnote{As for TNC, this action can be obtained as the dimensional reduction of the standard NS-NS sector of the supergravity actions, with (relativistic) metric given by $g_{\mu\nu}$ appearing in \eqref{NullCar}. Furthermore note that when $\Phi^{(Carroll)}=0$ this action correctly reduces to \eqref{TNCaction} with $\Phi^{(TNC)}=0$.}

\begin{align}\label{Carrollaction}
	\begin{split}
		S =& \int d^d x\, e\, \left[ \mathcal{R} + \frac{1}{2}a^\mu a_\mu + \frac{1}{2}\mathfrak{e}^\mu \mathfrak{e}_\mu-4a^\mu D_\mu \phi+4D^\mu\phi D_\mu \phi +2 \mathcal{K}  \upsilon^\mu D_\mu \Phi    +4 \upsilon^\mu D_\mu \phi \, \upsilon^\nu D_\nu \Phi \right.\\
		&\hphantom{(((\int d^d x e }   \left.  +2\Phi\left(\mathcal{K}^{\mu\nu}\mathcal{K}_{\mu\nu}-\mathcal{K}^2 -4\upsilon^\mu D_\mu \phi\,\upsilon^\nu D_\nu \phi -4\mathcal{K} \upsilon^\mu D_\mu \phi+\frac{1}{4}\upsilon^\rho \upsilon^\sigma H^{\mu\nu}_{\hphantom{\mu\nu}\rho}H_{\mu\nu\sigma}    \right)\right.  \\
		&\hphantom{(((\int d^d x e }   \left. -\frac{1}{12}H_{\mu\nu\rho}H^{\mu\nu\rho}+\frac{1}{2}b^{\mu\nu}H_{\mu\nu\rho}\upsilon^\rho   \right]
	\end{split}
\end{align}
with $\mathcal{K}\equiv  \hat h^{\mu\nu}\mathcal{K}_{\mu\nu}=-D_\mu \upsilon^\mu$,  $\mathcal{R}\equiv \hat h^{\mu\nu}\mathcal{R}_{\mu\nu}$ and the measure is defined as
\begin{equation}
e=e^{-2\phi}\sqrt{\frac{2\Phi}{ \det \bar h^{\mu\nu}}}=e^{-2\phi}\sqrt{2\Phi \det \bar h_{\mu\nu}},
\end{equation}
where $\bar h_{\mu\nu}=h_{\mu\nu}-\frac{\hat \tau_\mu \hat \tau_\nu}{2\Phi}$. We choose the independent fields of our theory to be $\Phi, \upsilon^\mu, \hat h^{\mu\nu}, B_{\mu\nu}$ and $\aleph_\mu$, see Appendix \ref{AppCar} for useful identites.
 
The variation of the action with respect to $\Phi$ gives 
\begin{equation}
\mathcal{R}_{\mu\nu}\upsilon^\mu \upsilon^\nu = \frac{1}{4}\upsilon^\rho \upsilon^\sigma H^{\mu\nu}_{\hphantom{\mu\nu}\rho}H_{\mu\nu\sigma}-2\upsilon^\mu \upsilon^\nu D_\mu D_\nu \phi .
\end{equation}
Note that contrary to what happened for TNC, this equation does not impose any constraint on the antisymmetric part of the connection, however it does impose a constraint on intrinsic torsion, which in the Carrollian case is given by the extrinsic curvature  \cite{Figueroa-OFarrill:2020gpr}.

The equations for the matter fields are
\begin{align}
D_\mu D^\mu \phi +a^\mu D_\mu\phi -2 D_\mu\phi  D^\mu \phi & = 2 \upsilon^\mu \upsilon^\nu \left(\Phi D_\mu  D_\nu \phi-2\Phi D_\mu \phi D_\nu \phi+ D_\mu \phi D_\nu \Phi \right)  -2 \mathcal{K}\Phi \upsilon^\mu D_\mu \phi \nonumber\\
&\quad  -\frac{1}{12}H^{\mu\nu\rho}H_{\mu\nu\rho}+\frac{1}{2}\upsilon^\rho \upsilon^\sigma \Phi H_{\rho}^{\ \mu\nu}H_{\sigma \mu\nu} +\frac{\mathfrak{e}^\mu \mathfrak{e}_\mu}{2}+\frac{1}{2} \upsilon^\rho b^{\mu\nu}H_{\rho\mu\nu}   \\
D^\rho b_{\rho\mu}-2 D^\rho \phi\, b_{\rho\mu}&=- 2\Phi\upsilon^\rho D_\rho \mathfrak{e}_\mu-2 \Phi\mathfrak{e}^\rho \mathcal{K}_{\rho\mu}+2 \Phi \mathfrak{e}_\mu \mathcal{K} +4\Phi \upsilon^\rho D_\rho \Phi \, \mathfrak{e}_\mu\nonumber   \\
&\quad +\frac{1}{2}F^{\rho\sigma}H_{ \rho \sigma\mu} +2D^\rho \Phi \, \upsilon^\sigma H_{\rho\sigma \mu}-2\Phi\upsilon^\rho a^\sigma H_{\rho\sigma\mu}\\
 D^\rho H_{\rho\mu\nu}+a^\rho H_{\rho\mu\nu}-2D^\rho\phi \,H_{\rho\mu\nu}&=\upsilon^\rho D_\rho b_{\mu\nu}-2\upsilon^\rho D_\rho \phi\, b_{\mu\nu}-2\mathcal{K}_{[\mu}^{\hphantom{\mu}\rho}b_{\nu]\rho} -\mathcal{K}b_{\mu\nu}  \nonumber\\
& \quad  +2\Phi \upsilon^\rho \upsilon^\sigma \left( D_\rho H_{\sigma\mu\nu}-2D_\rho \phi \, H_{\sigma\mu\nu}\right) +2\upsilon^\rho \upsilon^\sigma D_\rho \Phi \, H_{\sigma\mu\nu}\nonumber\\
&\quad+4\Phi\upsilon^\sigma \mathcal{K}_{[\mu}^{\hphantom{\mu}\rho}H_{\nu]\rho\sigma} -2 \mathcal{K} \Phi \upsilon^\rho H_{\rho\mu\nu} .
\end{align}

The equation for $\upsilon^\mu$ is
\begin{align}
\begin{split}
D^\rho F_{\rho\mu}+a^\rho F_{\rho\mu}-2 D^\rho \phi\, F_{\rho\mu}&= -4\Phi\upsilon^\rho D_{(\rho} a_{\mu)}+2 \Phi a_\mu \mathcal{K} +4\Phi \upsilon^\rho D_\rho \phi \, a_\mu-2 \upsilon^\rho D_\rho \Phi \, a_\mu \\
&\quad  +2 \upsilon^\rho D_\rho D_\mu \Phi+2\mathcal{K}_{\rho\mu}D^\rho \Phi -2\mathcal{K}D_\mu \Phi  -4 \upsilon^\rho D_\rho \phi\, D_\mu \Phi\\
&\quad +\frac{1}{2}b^{\rho\sigma}H_{\rho\sigma\mu}-\mathfrak{e}^\rho b_{\rho\mu}-2\Phi  \upsilon^\rho\mathfrak{e}^\sigma H_{\rho\sigma\mu} .
\end{split}
\end{align}

 Einstein's equations are given by
\begin{align} 
\begin{split}
 \mathcal{R}_{(\mu\nu)} +2 D_{(\mu}D_{\nu)}\phi-\frac{1}{4}H_{\mu}^{\hphantom{\mu}\rho\sigma}H_{\nu\rho\sigma}&=\frac{a_\mu a_\nu-\mathfrak{e}_\mu\mathfrak{e}_\nu}{2}+D_{(\mu}a_{\nu)} -F_{(\mu}^{\hphantom{\mu}\rho}\mathcal{K}_{\nu)\rho} +\upsilon^\rho b_{(\mu}^{\hphantom{\mu}\sigma}H_{\nu)\rho\sigma}  \\ 
&\quad+2\upsilon^\rho  \mathcal{K}_{\mu\nu}D_\rho \Phi -2 \Phi \mathcal{K}\, \mathcal{K}_{\mu\nu}+2\Phi\upsilon^\rho D_\rho \mathcal{K}_{\mu\nu} \\
 &\quad-4\Phi \upsilon^\rho\mathcal{K}_{\mu\nu} D_\rho \phi -\Phi \upsilon^\lambda \upsilon^\kappa H_{\mu\lambda}^{\hphantom{\mu\lambda}\rho}H_{\nu\kappa\rho}.
\end{split}
\end{align}
 
Once again one equation is missing, but it can be found directly from DFT. It is found to be 
\begin{align}
\begin{split}
D^\mu D_\mu \Phi -a^\mu D_\mu \Phi - 2D^\mu \phi D_\mu \Phi+(a^2-\mathfrak{e}^2)\Phi &=\frac{1}{4}\left(F^{\mu\nu}F_{\mu\nu}-b^{\mu\nu}b_{\mu\nu}\right)-2\mathcal{K}\Phi \upsilon^\rho D_\rho \Phi \\
&\quad +2\Phi \upsilon^\rho \upsilon^\sigma \left( D_\rho D_\sigma \Phi -2 D_\rho \phi D_\sigma \Phi \right).
\end{split}
\end{align}

It is also possible to rewrite these equations in terms of $\bar h^{\mu\nu}=\hat h^{\mu\nu}+2\Phi \upsilon^\mu \upsilon^\nu $ rather than $\hat h^{\mu\nu}$. The new connection $\bar \Gamma^\rho_{\mu\nu}$ is defined as
\begin{equation}
\Gamma^\rho_{\mu\nu}=\bar\Gamma^\rho_{\mu\nu}+2\Phi \upsilon^\rho \mathcal{K}_{\mu\nu}
\end{equation}
where quantities with a bar on them are understood to be defined with $\bar h^{\mu\nu}$ instead of $\hat h^{\mu\nu}$, i.e.
\begin{equation}
\bar\Gamma^\rho_{\mu\nu}=-\upsilon^\rho \partial_\mu \hat \tau_\nu +\frac{1}{2}\bar  h^{\mu\lambda}\left(\partial_\mu h_{\nu\lambda}+\partial_\nu h_{\mu\lambda}-\partial_\lambda h_{\mu\nu}\right) .
\end{equation}
The new Ricci tensor is related to the old one via
\begin{equation}
\mathcal{R}_{\mu\nu}={\mathcal{\bar R}}_{\mu\nu} - 2\upsilon^\rho  \bar D_\rho \Phi\, \mathcal{K}_{\mu\nu}+2\Phi \mathcal{K}\, \mathcal{K}_{\mu\nu}-2\Phi \upsilon^\rho \bar D_\rho \mathcal{K}_{\mu\nu} . 
\end{equation}
The action and equations will look nicer when using this connection, however $\bar h^{\mu\nu}, \hat h^{\mu\nu}$ and $\hat \tau_\mu$ are not compatible now:
\begin{align}
\begin{split}
\bar D_{\mu} \bar h^{\rho\sigma}&= 2 \upsilon^{\rho}\upsilon^\sigma \bar D_\mu \Phi -8 \mathcal{K}_{\mu\lambda} \bar h^{\lambda (\rho}\upsilon^{\sigma)}\Phi,\\
\bar D_\mu \hat h^{\rho\sigma}&=-4  \mathcal{K}_{\mu\lambda} \bar h^{\lambda (\rho}\upsilon^{\sigma)}\Phi,\\
\bar D_\mu \hat \tau_\nu &= -2\mathcal{K}_{\mu\nu}\Phi.
\end{split}
\end{align}

The action can then be rewritten as
\begin{align}\label{Carrollaction2}
	\begin{split}
		S =& \int d^d x\, e\, \left[ \mathcal{\bar R} + \frac{1}{2}a^\mu a_\mu + \frac{1}{2}\mathfrak{e}^\mu \mathfrak{e}_\mu-4a^\mu \bar D_\mu \phi+4\bar D^\mu\phi \bar D_\mu \phi +4 \upsilon^\mu \upsilon^\nu\bar  D_\mu \phi \,\bar D_\nu \Phi   \right.\\ 
		&\hphantom{(((\int d^d x e  }   \left.-8\Phi\mathcal{K} \upsilon^\rho \bar D_\rho \phi -\frac{1}{12}H_{\mu\nu\rho}H^{\mu\nu\rho}+\frac{1}{2}b^{\mu\nu}\upsilon^\rho H_{\mu\nu\rho}  \right],
	\end{split}
\end{align}
where now the tensors with upper indices are defined using $\bar h^{\mu\nu}$ and we will always write any expression such that no derivatives act on $\bar h^{\mu\nu}$, e.g. $\bar D^\mu \phi= \bar h^{\mu\nu}\bar D_{\nu} \phi \neq \bar  D_\nu \left(\bar h^{\mu\nu}\phi\right)$. The resulting equations of motion for the geometric fields are
\begin{align} 
\mathcal{\bar R}_{\mu\nu}\upsilon^\mu \upsilon^\nu &= \frac{1}{4}\upsilon^\rho \upsilon^\sigma H^{\mu\nu}_{\hphantom{\mu\nu}\rho}H_{\mu\nu\sigma}-2\upsilon^\mu \upsilon^\nu\bar D_\mu \bar D_\nu \phi  \label{CarGeo1}\\
\bar D^\rho F_{\rho\mu}+a^\rho F_{\rho\mu}-2\bar  D^\rho \phi\, F_{\rho\mu}&= \frac{1}{2}b^{\rho\sigma} H_{\rho\sigma\mu}+\mathfrak{e}^\rho b_{\mu\rho}+2\bar D^\rho \mathcal{K}_{\rho\mu}-2 \mathcal{K}\bar D_\mu \Phi \nonumber \\
&\quad +2 \upsilon^\rho \bar D_\mu \bar D_\rho \Phi -4 \upsilon^\rho \bar D_\rho \phi \, D_\mu \Phi \label{CarGeo2}\\
 \mathcal{\bar R}_{(\mu\nu)} +2\bar D_{(\mu}\bar D_{\nu)}\phi-\frac{1}{4}H_{\mu}^{\hphantom{\mu}\rho\sigma}H_{\nu\rho\sigma}&=\frac{a_\mu a_\nu-\mathfrak{e}_\mu\mathfrak{e}_\nu}{2}+D_{(\mu}a_{\nu)} - F_{(\mu}^{\hphantom{\mu}\rho}\mathcal{K}_{\nu)\rho}+\upsilon^\rho b_{(\mu}^{\hphantom{\mu}\sigma}H_{\nu)\rho\sigma} \nonumber\\
& \quad +4\Phi\upsilon^\rho \left( \bar D_\rho \mathcal{K}_{\mu\nu}-\mathcal{K}\,\mathcal{K}_{\mu\nu} +\frac{\bar D_\rho\Phi }{\Phi} \mathcal{K}_{\mu\nu} \right) \label{CarGeo3}
\end{align} 
 The equations for the matter fields are
\begin{align}
\bar D_\mu \bar D^\mu \phi +a^\mu \bar D_\mu\phi -2\bar  D_\mu\phi \bar  D^\mu \phi & = 2 \upsilon^\mu \upsilon^\nu D_\mu \Phi D_\nu \phi -\frac{1}{2}H^{\rho\mu\nu}H_{\rho\mu\nu}+\frac{1}{2}\mathfrak{e}^\mu\mathfrak{e}_\mu+\frac{1}{2} \upsilon^\rho b^{\mu\nu}H_{\rho\mu\nu}   \label{CarMat1}\\
\bar D^\rho b_{\rho\mu}-2 \bar D^\rho \phi\, b_{\rho\mu}&=  \frac{1}{2}F^{\rho\sigma}H_{\rho\sigma\mu}+2\upsilon^\rho \bar D^\sigma \Phi \, H_{\rho\sigma\mu}  \label{CarMat2}\\
 \bar D^\rho H_{\rho\mu\nu}+a^\rho H_{\rho\mu\nu}-2\bar D^\rho\phi \,H_{\rho\mu\nu}&=2b_{[\mu}^{\hphantom{[\mu}\rho}\mathcal{K}_{\nu]\rho}-\mathcal{K}b_{\mu\nu}+\upsilon^\rho \left(\bar D_\rho b_{\mu\nu} -2\bar D_\rho \phi\, b_{\mu\nu}+2\upsilon^\sigma \bar D_\sigma \Phi \, H_{\rho\mu\nu} \right).\label{CarMat3}
\end{align}  
 The "missing" equation is 
\begin{align}
\bar D^\mu \bar D_\nu \Phi -a^\mu \bar D_\mu \Phi -2 \bar D^\mu \phi \,\bar D_\mu \Phi &= \frac{1}{4}\left(F^{\mu\nu}F_{\mu\nu}-b^{\mu\nu}b_{\mu\nu}\right).
\end{align}

As in the case of TNC, we note that the equations \eqref{CarGeo1}-\eqref{CarMat3} are manifestly invariant under transformations corresponding to the Carroll algebra. We also recall the following definitions: 
\begin{align}
\begin{split}
a_{\mu} &= \upsilon^\rho F_{\rho \mu} = 2\upsilon^\rho\partial_{[\rho}\hat\tau_{\mu]},\\ 
\mathfrak{e}_{\mu} &= \upsilon^\rho b_{\rho \mu} = 2 \upsilon^\rho\partial_{[\rho}\aleph_{\mu]},\\
\mathcal{K}_{\mu\nu}&=-\frac{1}{2}\mathcal{L} _{\upsilon} h_{\mu\nu}=-\frac{1}{2}\left( \upsilon^\rho \partial_\rho h_{\mu\nu}+\left(\partial_\mu \upsilon^\rho\right)h_{\rho\nu}+\left(\partial_\nu \upsilon^\rho\right)h_{\mu\rho}\right)\, ,
\end{split}
\end{align}
which have the physical interpretation as the acceleration field, the electric and the extrinsic curvature of the geometry.
\section{String Newton-Cartan geometry } \label{SNCsec}
\subsection{Basics}
In   \cite{Seiberg:2000ms,Gopakumar:2000na,Klebanov:2000pp,Gomis:2000bd, Danielsson:2000gi} a non-relativistic string theory was formulated, which was then found to correspond to a target space geometry called String Newton-Cartan (SNC)  \cite{Gomis:2005pg, Brugues:2004an,Andringa:2012uz, Brugues:2006yd, Bergshoeff:2019pij,Bergshoeff:2018yvt}. A $D$-dimensional SNC spacetime naturally splits into two \textit{longitudinal} directions and $D-2$ \text{transverse} directions, that are mapped to each other by means of \textit{string} Galilean boosts. Recent works have delved deeper into the quantum aspects of such non-relativistic string theory, in  particular studying the Weyl symmetry and computing the beta functions, that are required to vanish for the theory to be anomaly-free  \cite{YanYu, Gomis:2019zyu}.

The basic geometric fields are  the longitudinal and transverse vielbeins, $\tau_\mu^A, \upsilon^\mu_A, E_\mu^{A'}, E^\mu_{A'}$ where $A$ runs over the two longitudinal directions and $A'$ runs over the remaining transverse directions. Using the transverse vielbeins we can build, as usual, two tensors $h^{\perp}_{\mu\nu}$ and $h^{\mu\nu}$. These fields satisfy the following  completeness relations
\begin{equation}
h^{\mu\rho}h^\perp_{\rho\nu}-\upsilon^\mu_A \tau_\nu^A=\delta^\mu_\nu,\qquad \qquad \upsilon^\mu_A \tau_\mu ^B=-\delta_A^B, \qquad \qquad \tau_\mu^A h^{\mu\rho}=\upsilon^\mu_A h^{\perp}_{\mu\rho}=0
\end{equation}
and transform under string boosts with parameter $\Sigma_A^{A'}$ as
\begin{equation}
\delta_\Sigma \upsilon^\mu_A =-E^\mu _{A'} \Sigma _A^{A'},\qquad \qquad \delta_\Sigma E_\mu^{A'} =-\tau_\mu^A \Sigma _A^{A'} , \qquad\qquad  \delta E^\mu_{A'}=\delta_{\Sigma} \tau_\mu ^A=0.
\end{equation}
The metric of the longitudinal space is $\eta_{AB}=\text{diag}\,   \left(-1,1\right)$ and our convention for the longitudinal Levi-Civita symbol is $\epsilon_{01}=+1$. 

Furthermore we can introduce a $Z_A$ symmetry in the theory via the gauge field $m_\mu^A$, which transforms under boosts as
\begin{equation}
\delta_{\Sigma} m_\mu^A = E^{A'}_\mu \Sigma_{A'}^A.
\end{equation}
This allows us to build the following manifestly boost invariant quantities:
\begin{align}
\begin{split}
\bar h_{\mu\nu} &= h^{\perp}_{\mu\nu} + 2\eta_{AB} m_{(\mu}^A\tau_{\nu)}^B,\\
u^\mu_A &= \upsilon^\mu_A + h^{\mu\rho} m_{\rho A},\\
\Phi^{AB}&= 2u^{\rho(A}m_{\rho}^{\ B)} - h^{\mu\nu}m_{\mu}^{A}m_{\nu}^{B}  = -u^{\mu A}u^{\nu B}\bar h_{\mu\nu},
\end{split}
\end{align}
which satisfy 
\begin{equation}
h^{\mu\rho}\bar h_{\rho\nu}-u^\mu_A \tau_\nu^A=\delta^\mu_\nu,\qquad \qquad u^\mu_A \tau_\mu ^B=-\delta_A^B, \qquad u^\mu_A \bar h_{\mu\rho}= \Phi_{A B} \tau_\rho^B
\end{equation}
The $Z_A$ transformations of the fields are
\begin{align}
\begin{split}
\delta_Z m_\mu^A &= D_\mu \sigma^A,\\
\delta_Z \bar h_{\mu\nu} &= 2\eta_{AB} \tau_{(\mu}^{AB} D_{\nu)} \sigma^B \\
\delta_Z u^\mu_A &= h^{\mu\rho}D_\rho \sigma^A\\
\delta_Z \Phi ^{AB} &= 2 u^{\rho (A}D_\rho \sigma^{B)}.
\end{split}
\end{align}
Note that this transformation is a symmetry of SNC only if the foliation constraint is imposed. However, one can avoid imposing this constraint by requiring the $B$-field to transform as well  \cite{Harmark:2019upf}, the full $Z_A$ transformations are then given by the upcoming \eqref{SNCZ}.

Using these fields it is also possible to construct the following boost-invariant connection:
\begin{align} \label{SNCcon}
\begin{split}
\Gamma^\rho_{\mu\nu}&= -u^\rho_A \left(\partial_\mu \tau_\nu^A+\omega_\mu \epsilon^A_{\ B}\tau_\nu^B\right)+\frac{1}{2}h^{\rho\sigma}\left(\partial_\mu \bar h_{\nu\sigma}+\partial_\nu \bar h_{\mu\sigma}-\partial_\sigma \bar h_{\mu\nu}\right)\\
&= -u^\rho_A\nabla_\mu \tau_\nu^A +\frac{1}{2}h^{\rho\sigma}\left(\partial_\mu \bar h_{\nu\sigma}+\partial_\nu \bar h_{\mu\sigma}-\partial_\sigma \bar h_{\mu\nu}\right)
\end{split}
\end{align}
where we introduced the spin connection $\omega_\mu^{\ AB}\equiv\omega_\mu \epsilon^{AB}$, associated with the longitudinal covariant derivative $\nabla_\mu$. The connection \eqref{SNCcon} is compatible with $h^{\mu\nu}$ and $\tau_\mu^A$ and has an antisymmetric component
\begin{equation}
2\Gamma^\rho_{[\mu\nu]} \equiv -u^{\rho}_A F_{\mu\nu}^A, 
\end{equation}
where we defined the  ``torsion" tensors as\footnote{Note that $F_{\mu\nu}^A=0$ is the so called foliation constraint.}
\begin{equation}
F_{\mu\nu}^A\equiv 2 \partial_{[\mu}\tau_{\nu]}^{\ A}+2\epsilon^A_{\ B} \omega_{[\mu}\tau_{\nu]}^{\ B}= 2 \nabla_{[\mu}\tau_{\nu]}^A.
\end{equation}
We can also decompose these tensors as 
\begin{equation}
F_{\mu\nu }^A = f^A \epsilon_{BC} \tau^B_\mu \tau^C_\nu +2a_{[\mu}^{\ \ BA}\tau_{\nu] B} +\widetilde{F}_{\mu\nu}^A
\end{equation}
where we defined the acceleration $a_{\mu AB}$, the temporal part of torsion $f^A$ and the transverse torsion tensor $\widetilde{F}_{\mu\nu}^A$ as 
\begin{align}
u^\rho_A F_{\rho\mu B}\equiv a_{\mu AB}\, ,\qquad u^{\mu A}a_\mu^{\ BC}=\epsilon^{AB}f^C\, ,\qquad 
u^{\mu A}\widetilde{F}_{\mu\nu B}=0\, .
\end{align}

Finally we define the extrinsic curvature in the usual way:
\begin{equation}
\mathcal{K}_{\mu\nu}^{A} =-\frac{1}{2} \mathcal{L}_{u } \bar h_{\mu\nu}
\end{equation}
which satisfies in particular 
\begin{align}
\begin{split}
\mathcal{K}_{\mu\nu}^A h^{\mu\nu} &= -D_\mu u^{\mu A} ,\\
h^{\mu\nu}h^{\rho\sigma} \mathcal{K}_{\mu\rho}^A\mathcal{K}_{\nu\sigma}^B&=D_\mu u^{\nu A}D_\nu u^{\mu B}+\frac{1}{4} h^{\mu\nu}h^{\rho\sigma}F_{\mu\nu C}F_{\rho\sigma D}\Phi ^{AC}\Phi^{BD}.
\end{split}
\end{align}
Below we discuss how this structure is embedded in the double field theory framework. 
\subsection{Embedding in DFT} 

To find the embedding of SNC in DFT we will use the same approach used for TNC  \cite{Blair:2019qwi}, namely we will compare the SNC worldsheet action  \cite{Gomis:2019zyu},
\begin{align} \label{SNCWS}
\begin{split}
S_{SNC}=& -\frac{1}{2}\int d^2\sigma \left[ e\, \partial X^\mu \partial X^\nu\, \bar h_{\mu\nu} +\epsilon^{\alpha\beta} \left(\lambda e_\alpha \tau_\mu +\bar \lambda \bar e_\alpha \bar \tau_\mu \right) \partial_\beta X^\mu\right]\\
&-\frac{1}{2}\int d^2\sigma \,   \varepsilon^{\alpha\beta} \partial_\alpha X^\mu \partial_\beta X^\nu B_{\mu\nu} ,
\end{split}
\end{align}
with the worldsheet action written in terms of DFT fields\footnote{We define $e_\alpha\equiv e^0_\alpha +e^1_\alpha$, $\bar e_\alpha \equiv e^0_\alpha -e^1_\alpha$, and similarly for $\tau$ and $\bar \tau$. Notice that $\gamma^{\alpha\beta}=\eta^{ab}e^\alpha_a e^\beta_b =-e^\alpha_0 e^\beta_0+e^\alpha_1 e^\beta_1 = -4 e^{(\alpha}\bar e^{\beta)}$ where $e^\alpha \equiv \frac{1}{2}(e^\alpha_0+e^\alpha_1 )$, $\bar e^\alpha \equiv \frac{1}{2}(e^\alpha_0-e^\alpha_1 )$, such that $e^\alpha e_\alpha= 1= \bar e^\alpha \bar e_\alpha$ and $e^\alpha \bar e_\alpha=0$. Moreover we have $\varepsilon^{\alpha\beta}= e \epsilon^{ab}e^\alpha_a e^\beta _b$ with $\varepsilon^{01}=-\varepsilon_{01}=+1$ and $e^\alpha_a= e^{-1} \varepsilon^{\alpha\beta} e_\beta ^b \varepsilon_{ba}$, which implies $e^\alpha = -\frac{1}{2e} \varepsilon^{\alpha\beta}\bar e_\beta$ and  $\bar e^\alpha = \frac{1}{2e} \varepsilon^{\alpha\beta} e_\beta$. Note that our convention for the worldsheet Levi-Civita symbol $\varepsilon^{\alpha\beta}$ is the opposite of the one for the longitudinal SNC Levi-Civita symbol $\epsilon^{AB}$.   },
\begin{align}
\begin{split}
S_{DFT}=& -\frac{1}{2}\int d^2\sigma \left[ e\, \partial X^\mu \partial X^\nu\, K_{\mu\nu}  + 2\epsilon^{\alpha\beta} \left(\beta_{\alpha a}x^a_\mu +\bar \beta_{\alpha\bar a}\bar x^{\bar a}_\mu\right) \partial_\beta X^\mu\right]\\
&-\frac{1}{2}\int d^2\sigma \left[  \varepsilon^{\alpha\beta} \partial_\alpha X^\mu \partial_\beta X^\nu \mathcal{B}_{\mu\nu} -2\gamma^{\alpha\beta} \left(\beta_{\alpha a}x^a_\mu -\bar \beta_{\alpha\bar a}\bar x^{\bar a}_\mu\right) \partial_\beta X^\mu \right].
\end{split}
\end{align}
To make the identifications needed to define the SNC embedding we need to rewrite \eqref{SNCWS} in a more suitable way. In particular there are two issues we need to address:
\begin{enumerate}
\item We need to have a term proportional to $\gamma^{\alpha\beta}\lambda\partial_{\beta} X^\mu$ and $\gamma^{\alpha\beta}\bar \lambda\partial_{\beta} X^\mu$ to be able to properly identity the null eigenvectors of DFT;
\item We need to make sure that $K_{\mu\nu}$ is a degenerate matrix, which is not the case if we naively identify $K_{\mu\nu}=\bar h_{\mu\nu}$.
\end{enumerate}
The first issue is simply solved by the following field redefinition:
\be
A_\alpha= \frac{1}{2}(\lambda - \bar \lambda)e^0_\alpha +\frac{1}{2}(\lambda + \bar \lambda)e^1_\alpha = \frac{1}{2} \lambda e_\alpha -\frac{1}{2}\bar \lambda \bar e_\alpha.
\ee 
 In terms of these new Lagrange multipliers the SNC action becomes
\begin{align}
\begin{split}\label{SNCWS2}
S_{SNC}=& -\frac{1}{2}\int d^2\sigma \left[ e\, \partial X^\mu \partial X^\nu\, \bar h_{\mu\nu} +2\epsilon^{\alpha\beta}\tau_\mu^1 A_\alpha\partial_\beta X^\mu\right]\\
&-\frac{1}{2}\int d^2\sigma \left[ \epsilon^{\alpha\beta} \partial_\alpha X^\mu \partial_\beta X^\nu B_{\mu\nu}+2e\gamma^{\alpha\beta}\tau_\mu^0 A_\alpha\partial_\beta X^\mu\right] ,
\end{split}
\end{align} 
This form of the action makes it easy for us to identify the null eigenvectors of DFT, however we still have to solve problem (2), i.e. the fact that $K_{\mu\nu}=\bar h_{\mu\nu}$ is not a degenerate matrix. To solve this issue we can make use of the Stueckelberg symmetry of the SNC action. The action \eqref{SNCWS} is invariant under the following transformations with parameters $C^A$  \cite{Harmark:2019upf, Bergshoeff:2019pij}:
\begin{align}
\delta \bar h_{\mu\nu}& =2 C_{(\mu}^A \tau_{\nu)}^B \eta_{AB} ,& \delta B_{\mu\nu} &=-2C_{[\mu}^A \tau_{\nu]}^B \epsilon_{AB} \\
\lambda' &= \lambda + e^{-1} \epsilon^{\alpha \beta}\bar e_\alpha \bar C_\beta,&\bar \lambda' &=\bar \lambda +  e^{-1} \epsilon^{\alpha\beta} e_\alpha C_\beta,
\end{align}
where $C_\mu=C^0_\mu+C^1_\mu$ and $\bar C_\mu=C_\mu^0-C_\mu^1$. This means we can rewrite the action in terms of $h^{\perp}_{\mu\nu}$ and $\bar B_{\mu\nu}$ by choosing $C_\mu^A = -m_\mu^A$. After redefining the Lagrange multipliers one more time we arrive at the action 
\begin{align}
\begin{split}\label{SNCWS3}
S_{SNC}=& -\frac{1}{2}\int d^2\sigma \left[ e\, \partial X^\mu \partial X^\nu\, h^{\perp}_{\mu\nu} +2\epsilon^{\alpha\beta}\tau_\mu^1 A_\alpha\partial_\beta X^\mu\right]\\
&-\frac{1}{2}\int d^2\sigma \left[ \epsilon^{\alpha\beta} \partial_\alpha X^\mu \partial_\beta X^\nu \bar B_{\mu\nu}+2e\gamma^{\alpha\beta}\tau_\mu^0 A_\alpha\partial_\beta X^\mu\right] ,
\end{split}
\end{align} 
where now all the tensors appearing in \eqref{SNCWS3} are manifestly $Z_A$ invariant. This allows us to make the following identifications:
\be \label{KBid}
K_{\mu\nu}=h^\perp_{\mu\nu},\qquad \qquad \mathcal{B}_{\mu\nu}=\bar B_{\mu\nu}
\ee
and 
\begin{align}
\begin{split}
  \left(\beta_{\alpha }x _\mu +\bar \beta_{\alpha}\bar x_\mu\right)& = \left(\tau_\mu +\bar \tau_\mu \right) A_\alpha=2\tau_\mu^0  A_\alpha\\
 \left(\beta_{\alpha }x_\mu -\bar \beta_{\alpha }\bar x_\mu\right) &=-\left(\tau_\mu -\bar \tau_\mu \right) A_\alpha=-2 \tau_\mu^1 A_\alpha,
\end{split}
\end{align}
This is solved by (note the mismatch of bars between DFT and SNC fields)
\begin{align}
\begin{split}
x _\mu& = \frac{1}{\sqrt{2}}  \bar \tau_\mu  , \qquad \qquad \beta_{\alpha} = \frac{1}{\sqrt{2}} A_\alpha  \\
\bar x_\mu& =  \frac{1}{\sqrt{2}}    \tau_\mu  ,\qquad  \qquad \bar \beta_{\alpha } = \frac{1}{\sqrt{2}}  A_\alpha .
\end{split}
\end{align}    
By requiring $x_\mu y^\mu =1= \bar x_\mu \bar y^\mu $   we find
\be
y^\mu_a =-\sqrt{2}\,\bar \upsilon^\mu  , \qquad \qquad \bar y^\mu_{\bar a} = - \sqrt{2}\,\upsilon^\mu .
\ee 
Moreover we have
\be
H^{\mu\nu}= h^{\mu\nu}
\ee
and    
\be
Z_\mu^\nu = \tau_\mu \upsilon^\nu  - \bar \tau_\mu \bar \upsilon^\nu   .
\ee 

In summary, string Newton-Cartan can be embedded in DFT as
\begin{align}
K_{\mu\nu} & =h^\perp_{\mu\nu}, & H^{\mu\nu}&= h^{\mu\nu} \\
\mathcal{B}_{\mu\nu} &= B_{\mu\nu}, & Z_\mu^\nu &= -\tau_\mu \upsilon^\nu +\bar \tau_\mu  \bar \upsilon^\nu 
\end{align} 
 and the eigenvectors are
\begin{align}
\begin{split}
x _\mu& = \frac{1}{\sqrt{2}}\bar \tau_\mu  , \qquad\qquad  y^\mu =-\sqrt{2}\,\bar\upsilon^\mu   \\
\bar x _\mu& =  \frac{1}{\sqrt{2}} \tau_\mu  , \qquad\qquad \bar  y^\mu =- \sqrt{2}\,\upsilon^\mu .
\end{split}
\end{align}  
An issue with this parametrization is that it does not include the gauge field $m_\mu^A$. To reinstate it we can once again make a Stueckelberg transformation (or a shift transformation \eqref{shiftDFT}  from the point of view of DFT)
\begin{align} \label{DFTShift}
\begin{split}
\left(y^\mu\right)' &= y^\mu +H^{\mu\rho} V_\rho\\
\left(\bar y^\mu\right)' &= 	\bar y^\mu +H^{\mu\rho} \bar V_\rho \\
\left(K_{\mu\nu}\right)' &= K_{\mu\nu}-2x_{(\mu}K_{\nu)\rho}  V^\rho-2\bar x_{(\mu}K_{\nu)\rho}\bar V^\rho +\left(x_\mu  V_\rho +\bar x_\mu \bar V_\rho \right)\left(x_\nu  V^\rho +\bar x_\nu\bar V^\rho \right)  \\
\left(\mathcal{B}_{\mu\nu}\right)' &= \mathcal{B}_{\mu\nu}-2x_{[\mu}  V_{\nu]}+2\bar x_{[\mu}\bar  V_{\nu]} + 2x_{[\mu}\bar x_{\nu]}\left(y^\rho \bar V_\rho +\bar y^\rho   V_\rho +V_\rho \bar V^\rho \right)
\end{split}
\end{align}
where $V_\mu, \bar V_\mu$ are two arbitrary local parameters and we set $V^\rho \equiv H^{\rho\mu}V_\mu$ for brevity. The choice $V_\mu=\frac{1}{2}m_\mu, \bar V_{\mu}=\frac{1}{2}\bar m_{\mu}$ gives 
\begin{align}
\begin{split}
\left(y^\mu\right)' &= -\bar \upsilon^\mu +\frac{1}{2}  m^\mu\\
\left(\bar y^\mu\right)' &= 	 - \upsilon^\mu+\frac{1}{2}\bar   m^\mu \\
\left(K_{\mu\nu}\right)' &=h_{\mu\nu}-\frac{1}{2}\bar\varphi \tau_\mu \tau_\nu -\frac{1}{2}  \varphi \bar \tau_\mu \bar \tau_\nu -\mathcal{T} \tau_{(\mu}\bar \tau_{\nu)}  \\
\left(\mathcal{B}_{\mu\nu}\right)' &=B_{\mu\nu}+\mathcal{T}\tau_{[\mu}\bar \tau_{\nu]}
\end{split}
\end{align}
with 
\begin{align}\label{Newts}
\begin{split}
\varphi &\equiv 2\bar u^\mu m_\mu +\frac{1}{2}m^2 \\
\bar \varphi &\equiv 2u^\mu \bar m_\mu +\frac{1}{2}\bar m^2 \\
\mathcal{T} &\equiv  u^\mu m_\mu+\bar u^\mu \bar m_\mu +\frac{1}{2}m\cdot \bar m 
\end{split}
\end{align}
and
\begin{align} 
 u^\mu&\equiv \upsilon^\mu -\frac{1}{2}h^{\mu\rho}\bar m_{\rho}, & \bar u^\mu&\equiv \bar \upsilon^\mu -\frac{1}{2}h^{\mu\rho}m_{\rho}\\
\bar h_{\mu\nu}&=h^{\perp}_{\mu\nu}-\tau_{(\mu}\bar m_{\nu)}-\bar \tau_{(\mu} m_{\nu)},& B_{\mu\nu}&=\bar B_{\mu\nu}+\tau_{[\mu}\bar m_{\nu]}-\bar \tau_{[\mu} m_{\nu]}. 
\end{align}

Changing back from lightcone coordinates we finally find 
\begin{align}
\begin{split} \label{SNCDFT1}
K_{\mu\nu}&= \bar h_{\mu\nu}+ \Phi_{AB}\tau_{\mu}^A\tau_\nu^B,\\
H^{\mu\nu}&= h^{\mu\nu},\\
Z^\mu_\nu &= -\epsilon_{AB} u^{\mu A}\tau_\nu^B,\\
\mathcal{B}_{\mu\nu} &= B_{\mu\nu}-\frac{1}{2} \Phi \epsilon_{AB} \tau_\mu^A \tau_\nu^B,
\end{split}
\end{align}
where we defined $\Phi \equiv \Phi^A_A$.

\subsection{Action and equations of motion}

Given an arbitrary tensor $A_{\mu\nu\dots}$ we will define 
\begin{equation} \label{UpIndSNC}
A^{\mu\nu\dots}\equiv A_{\rho\sigma\dots} h^{\mu\rho}h^{\nu\sigma}\dots
\end{equation} 
The only SNC fields which naturally have upper \textit{curved} indices are $  u^\mu_A$ and   $h^{\mu\nu}$ hence any other tensor with upper indices is to be understood as defined via \eqref{UpIndSNC}.

Using the parametrization \eqref{SNCDFT1} we find that the SNC action is given by\footnote{Close to the completion of this work we became aware of the work of Bergshoeff et al. \cite{Bergshoeff:2020} where they obtained a similar action by taking a particular non-relativistic limit of NS-NS gravity.}
\begin{align}\label{SNCaction}
	\begin{split}
		S =& \int d^D x\, e\, \left[ \mathcal{R} -a^{\mu AB} (a_{\mu (AB)}  -\frac{1}{2}\eta_{AB} a_\mu )+\left( a^\mu -2D^\mu \phi \right)\left( a_\mu -2D_\mu \phi \right) \right.\\
		&\hphantom{(((\int d^d x e\, }   \left.  -\frac{1}{2} F^{\mu\nu A}F_{\mu\nu}^{B} (\Phi_{AB}-\frac{1}{2} \eta_{AB}\Phi )  +\frac{1}{2}\epsilon_{AB} u^{\rho A}F^{\mu\nu B}H_{\rho\mu\nu}   -\frac{1}{12}H^{\rho\mu\nu}H_{\rho\mu\nu}    \right]  \\ 
	\end{split}
\end{align}
where we defined $a_\mu \equiv  a_{\mu A}^{\ \ A}$,  $\Phi\equiv \Phi^A_A$ and the invariant measure is given by
\begin{equation}\label{SNCmeasure}
e \equiv  e^{-2\phi}\sqrt{\frac{\det \bar h}{\det \Phi}}.
\end{equation}
This action is invariant under the $Z_A$ transformations
\begin{align}
\begin{split}\label{SNCZ}
		\delta \bar h_{\mu\nu}=2 \tau_{(\mu}^{\ A}D_{\nu)}\sigma_A ,\qquad \qquad \delta B_{\mu\nu}=2\epsilon_{AB}\tau_{[\mu}^A D_{\nu]}\sigma^B ,\\
 \delta u^{\mu A} = h^{\mu\nu}D_\nu \sigma^A, \qquad \qquad   \delta \Phi^{AB}= 2 u^{\mu (A}D_\mu \sigma^{B)}. 
\end{split}
\end{align}

The equations of motion can be found by varying the action with respect to the independent fields $h^{\mu\nu}, u^\mu_A$ and $\Phi_{AB}$\footnote{Note that, despite appearances,  the action does not depend on the spin connection, i.e. $\delta S/\delta \omega_\mu =0$.}, also see Appendix \ref{AppSNC} for useful identities. Sometimes it will be useful to further decompose the acceleration in its antisymmetric and traceless symmetric components:
\begin{equation} \label{SNCaDecomp}
a_\mu^{\ AB}= \mathcal{S}_\mu^{AB} + \frac{1}{2}\eta^{AB} a_\mu + \epsilon^{AB}\mathcal{A}_\mu
\end{equation}
with $\mathcal{S}_\mu^{\ AB}\eta_{AB}=0$ and we recall $a_\mu=a_\mu^{\ AB}\eta_{AB}$.

Taking the variation of \eqref{SNCaction} with respect to $\Phi_{AB}$ we find the equation\footnote{The action actually only depends on the \textit{trace} of the tensor $\Phi_{AB}$, as can be seen by expanding the Ricci scalar and noticing that the $F^{\mu\nu A}F_{\mu\nu}^B \Phi_{AB}$ terms cancel out. This is why the corresponding equation is a singlet and not a tensor.}
\begin{equation}
F^{\mu\nu A}F_{\mu\nu A}=0,
\end{equation}
which we will use when writing the remaining equations.
The space projection of the equation of motion for $u^{\mu}_A$ is
\begin{align}\label{FssSNC}
D^\rho F^{\mu}_{\  \rho A}+a_{\rho A}^{\ \ B}F^{\mu\rho}_{\ \ B}-2 F^{\mu}_{\ \rho A}D^\rho \phi&= \frac{1}{2}\epsilon_{AB} F^{\rho\sigma B}H^{\mu}_{\ \rho\sigma} ,
\end{align}
while the time projection is 
\begin{align}
D^\rho \mathcal{S}_\rho^{\ AB}+ a^{\rho} \mathcal{S}_{\rho}^{\ AB}-2 \mathcal{S}^{\rho AB}D_\rho \phi &= -\frac{1}{4}F^{\mu\nu A}F_{\mu\nu}^{ B}\Phi + 2\epsilon^{(A}_{\ \ C}\mathcal{F}^{B)C} ,
\end{align}
where we defined
\begin{align}
\mathcal{F}^{BC}&\equiv -\mathcal{A}^\rho \mathcal{S}_\rho^{\ BC}+\frac{1}{4} u^{\rho (B}F_{\mu\nu}^{C)}H_{\rho}^{\ \mu\nu} . 
\end{align} 

The equation for the B-field is
\begin{align}
D^\rho H_{\rho}^{\ \mu\nu} + a^\rho H_{\rho}^{\ \mu\nu} - 2D^\rho \phi \, H_{\rho}^{\ \mu\nu}  &= 2 \Omega^{\mu\nu}+ \epsilon_{AB} \mathcal{H}^{\mu\nu AB},
\end{align}
where we defined
\begin{align}
\begin{split}
 \mathcal{H}^{\mu\nu AB}&\equiv 2 u^{\rho A}D_\rho F^{\mu\nu B}-2 u^{\rho A} F^{\mu\nu B} D_\rho \phi  -2D^{[\mu} a^{\nu] AB}    -2 \mathcal{S}^{\mu A}_{\ \ C}\mathcal{S}^{\nu BC}\\
&\quad  -\mathcal{K}^A  F^{\mu\nu B}+2  F^{[\mu}_{\ \ \rho C}F^{\nu]\rho A}\Phi^{BC} 
\end{split}
\end{align}
as well as $\Omega_{\mu\nu} \equiv 2 \partial_{[\mu} \omega_{\nu]}$ and $\mathcal{K}^A \equiv h^{\mu\nu}\mathcal{K}_{\mu\nu}^{A}$.

The dilaton equation is
\begin{align}
D^\mu D_\mu \phi -\frac{1}{2} D^\mu a_\mu  + 2 a_\mu D^\mu \phi -2 D^\mu \phi \, D_\mu \phi -\frac{1}{2} a_\mu a^\mu &= \frac{1}{4}\epsilon_{AB} u^{\rho A}F^{\mu\nu B}H_{\rho\mu\nu}-\frac{1}{12} H^{\rho\mu\nu}H_{\rho\mu\nu}.
\end{align}

The space-space projection of Einstein's equations is
\begin{align}
\begin{split}
\mathcal{R}^{(\mu\nu)} + 2 D^{(\mu}D^{\nu)}\phi -\frac{1}{4}H^\mu_{\ \rho\sigma}H^{\nu \rho\sigma} &= u^{\rho }_A D^{(\mu} F_\rho^{\ \nu)A}+\mathcal{S}^{\mu AB}\mathcal{S}^\nu_{\  AB} \\
&\ \ \    -\epsilon^{A B}u^{\sigma }_{\ A}F^{\rho (\mu}_{\ \ \  B}H^{\nu)}_{\ \ \rho \sigma} -\frac{1}{2} F^{\mu\rho A}F^\nu_{\ \rho A} \Phi . 
\end{split}
\end{align}
The time-space projection is
\begin{align}
\begin{split}
h^{\mu \rho}u^{\sigma A}\mathcal{R}_{(\rho\sigma)} + & 2 u^{\rho A}D^{\mu}D_{\rho}\phi +a^\mu u^{\rho A}D_\rho \phi -\frac{1}{4}H^\mu_{\ \rho\sigma}H_{\lambda}^{\  \rho\sigma}u^{\lambda A} = \frac{1}{2}u^{\rho A}u^{\sigma }_{\ B}D_\rho F_{\sigma}^{\ \mu B}-u^{\rho }_{\ B}D^\mu a_{\rho}^{\ (AB)}+u^{\rho A}D^\mu a_{\rho} \\
& -\frac{1}{2}u^{\rho B} D_{\rho}a^\mu_{BA}+\frac{1}{2}u^{\rho A} D_{\rho}a^\mu - 2a^{\mu [AB]}u^\rho_{\ B} D_\rho \phi-a^\mu u^{\rho A}D_\rho \phi   -\frac{1}{2} a^\mu \mathcal{K}^A+\frac{1}{2} a^{\mu BA}\mathcal{K}_B \\
& -F^{\mu}_{\ \rho C}a ^{\rho B(A}\Phi_{B}^{\ C)}  +a^{\rho}F^{\mu}_{\ \rho C}\Phi^{AC}+\frac{1}{2}a^\rho F_\rho^{\ \mu A}\Phi +\frac{1}{2}F^{\mu\rho B}D_\rho \Phi^A_{\ B} -\frac{1}{2}F^{\mu\rho A}D_\rho \Phi\\
&   + \epsilon^{AB}\mathcal{P}^\mu_{\ B},
\end{split}
\end{align}
where
\begin{align}
\begin{split}
\mathcal{P}^{\mu  B}&=- \frac{1}{2}u^{\sigma B}D^\rho H^\mu_{\ \rho\sigma} +u^{\rho B}H^{\mu}_{\ \rho \sigma}D^\sigma \phi +a^\rho u^{\lambda B}H^\mu_{\      \rho\lambda} +\frac{1}{2}F^{\mu \rho C}u^{\sigma B}u^{\lambda }_C H_{\rho \sigma \lambda} \\
&\quad-\frac{1}{2} a^{\rho CB}u^{\sigma}_{\ C} H^\mu_{\ \rho\sigma}  -D^\mu f^B -\frac{1}{2} a^\mu f^B + a^{\mu BC}f_C+\frac{1}{2} a^{\mu CB}f_C + a^{\mu B C} f_{C}
\end{split} 
\end{align}
Finally the time-time component (found directly from DFT) is given by
\begin{align}
\begin{split}
u^{\mu A}u^\nu_{\ A}\mathcal{R}_{\mu\nu}+&2 u^{\mu A}u^\nu_{\ A}D_\mu D_\nu \phi - \frac{1}{4}u^{\mu A}u^\nu_{\ A}H_{\mu\rho\sigma}H_\nu^{\ \rho\sigma} = \frac{3}{2}\mathcal{S}^{\rho AB}D_\rho \Phi _{AB}-\frac{3}{4}a^\mu D_\mu \Phi   \\
&-\frac{3}{4}a^\mu a_\mu \Phi + \mathcal{S}^{\mu AB}\mathcal{S}_{\mu A C}\Phi _B^{\ C}-2 f^A f_A +\epsilon^{AB}\mathcal{Q}_{AB}
\end{split}
\end{align}
where
\begin{align}
\begin{split} \label{QttSNC}
\mathcal{Q}_{AB}&=-\frac{1}{2} u^{\rho}_{\ A} u^\sigma_{\ B} D^\lambda H_{\rho\sigma \lambda} + u^\mu_A u^\nu_B H_{\mu\nu\rho} D^\rho \phi +u^\mu_A u^\nu_B  D_\mu \mathcal{A}_\nu +2 f_A u^\mu_B D_\mu \phi+\frac{3}{2}u^\mu_A D_\mu f_B\\
&\quad -u^\mu_A u^{\nu C}\mathcal{S}^{\rho}_{\ C B} H_{\mu\nu\rho} - u^\mu_A u^{\nu}_B a^\rho H_{\mu\nu\rho}+f_A \mathcal{K}_B + 3 \mathcal{A}^\mu \mathcal{S}_{\mu A C}\Phi_{B}^{\ C}\\
&\quad -\frac{1}{4}F^{\mu\nu }_{\ \ A} u^{\rho C}H_{\rho\mu\nu} \Phi_{BC}-\frac{1}{4}F^{\mu\nu }_{\ \ B} u^{\rho}_A H_{\rho\mu\nu} \Phi +\frac{1}{4}F^{\mu\nu C} u^{\rho}_A H_{\rho\mu\nu} \Phi_{BC}
\end{split}
\end{align}
We remind the reader of the definitions of some of the fields appearing in these equations:
\begin{align}
\begin{split}
F_{\mu\nu}^A &= 2\nabla_{[\mu}\tau_{\nu]}^A \equiv f^A \epsilon_{BC} \tau^B_\mu \tau^C_\nu +2a_{[\mu}^{BA}\tau_{\nu] B} +\widetilde{F}_{\mu\nu}^A
\\
a_{\mu}^{AB} &= u^{\rho A}F_{\rho\mu}^B\equiv  \mathcal{S}_\mu^{AB} + \frac{1}{2}\eta^{AB} a_\mu + \epsilon^{AB}\mathcal{A}_\mu.
\end{split}
\end{align}

These equations are clearly more complicated than the ones found for TNC or Carroll, as there is no obvious way to combine the time and space projections of an equation into a single simpler equation. Despite this technical difficulty, it is possible to show that these equations are invariant under the SNC algebra as expected. In the next section we will  discuss these equations and their relations to previous results .

\section{Comparison with known results} \label{CompSec}

\subsection{TNC beta functions}
\label{TNCbeta}
In  \cite{Gallegos:2019icg}, three of us computed the beta functions for string theory describing a Type I TNC target space.  The equations obtained by setting those beta functions to zero were the two twistlessness contraints
\begin{align}
	   \label{beta1} F_{\mu\nu}&= a_\mu \tau_\nu - \tau_\mu a_\nu \, , \\
 		\label{beta2} b_{\mu\nu}&= \mathfrak{e}_\mu \tau_\nu - \tau_\mu \mathfrak{e}_\nu \, ,
\end{align}
two scalar equations
\begin{align}
	\label{beta3}  D\cdot a + a^2 &=  2 \mathfrak{e}^2 + 2 \left(a \cdot D\phi \right) \, ,\\
 \label{beta4} D\cdot \mathfrak{e} &= 2 \left(\mathfrak{e} \cdot D\phi\right)  \, , 
\end{align}
and two tensor equations	
\begin{align}
  \label{beta5}  R_{(\mu\nu)} - \frac{1}{4}H^{\rho\sigma}_{\ \ (\mu} H_{\nu)\rho\sigma} + 2 D_{(\mu} D_{\nu)} \phi &= \frac{\mathfrak{e}^2 \left(2 \Phi \tau_\mu \tau_\nu - \bar h_{\mu\nu} \right) - \mathfrak{e}_\mu \mathfrak{e}_\nu}{2}   - \mathfrak{e}^\sigma \left( \Delta_T \right)^\rho_{(\mu} H_{\nu)\rho\sigma}   \nonumber \\
  &\hphantom{=} +\left(\Delta_S \right)^\rho_{(\mu} D_{\nu)}a_{\rho} + \frac{a_\mu a_\nu -2a^2\,\Phi \tau_\mu\tau_\nu}{2}     \, , \\
 \label{beta6} D^\rho H_{\rho \mu\nu}+a^\rho H_{\rho \mu\nu}   -2 H^\rho_{\ \mu\nu}D_\rho\phi &=2 \left(\Delta_S \right)^\rho_{[\mu} D_{\nu]} \mathfrak{e}_\rho -2\left( \Delta_T\right)^\rho_{[\mu} D_{\nu]} \mathfrak{e}_\rho  -2a_{[\mu} \mathfrak{e}_{\nu]}
   \nonumber  \\  &\hphantom{=} + \left( 2\hat \upsilon^\rho D_\rho \phi - D_\rho \upsilon^\rho \right)b_{\mu\nu} \, 
 \end{align}
where we remind the reader of the definitions of the projectors $\left(\Delta_T\right)^\mu_\nu \equiv -\hat \upsilon^\mu \tau_\nu$ and $\left(\Delta_S\right)^\mu_\nu \equiv h^{\mu\rho} \bar h_{\rho\nu}$ which satisfy
\be
\left(\Delta_S\right)^\mu_\nu + \left(\Delta_T\right)^\mu_\nu=\delta^\mu_\nu\, .
\ee 
We now want to compare these equations with the ones obtained from DFT. First of all, notice that in  \cite{Gallegos:2019icg}  twistless torsion was assumed from the start which explains the difference between the equations \eqref{beta1}-\eqref{beta2} and \eqref{twistTNC}. To compare the rest of the equations, we impose the twistlessness constraints in equations \eqref{MatterTNC}-\eqref{FTNC} and \eqref{EinsteinTNC}.  We then find two scalar equations 
 \begin{align}
	\label{beta3DFT}  D\cdot a + a^2 &=   \mathfrak{e}^2 + 2 \left(a \cdot D\phi \right) \, ,\\
 \label{beta4DFT} D\cdot \mathfrak{e} &= 2 \left(\mathfrak{e} \cdot D\phi\right)  \, , 
\end{align}
and two tensor equations	
\begin{align}
  \label{beta5DFT}   R_{(\mu\nu)} - \frac{1}{4}H^{\rho\sigma}_{\ \ (\mu} H_{\nu)\rho\sigma} + 2 D_{(\mu} D_{\nu)} \phi &= \frac{2\mathfrak{e}^2 \Phi \tau_\mu \tau_\nu - \mathfrak{e}_\mu \mathfrak{e}_\nu}{2}-\frac{2a^2\,\Phi \tau_\mu\tau_\nu-a_\mu a_\nu }{2}      \nonumber \\
  &\hphantom{=} +\left(\Delta_S \right)^\rho_{(\mu} D_{\nu)}a_{\rho}  - \mathfrak{e}^\sigma \left( \Delta_T \right)^\rho_{(\mu} H_{\nu)\rho\sigma}   \, , \\
 \label{beta6DFT} D^\rho H_{\rho \mu\nu}+a^\rho H_{\rho \mu\nu}   -2 H^\rho_{\ \mu\nu}D_\rho\phi &=2 \left(\Delta_S \right)^\rho_{[\mu} D_{\nu]} \mathfrak{e}_\rho -2\left( \Delta_T\right)^\rho_{[\mu} D_{\nu]} \mathfrak{e}_\rho  -2a_{[\mu} \mathfrak{e}_{\nu]}
   \nonumber  \\  &\hphantom{=} + \left( 2\hat \upsilon^\rho D_\rho \phi - D_\rho \upsilon^\rho \right)b_{\mu\nu} \, 
 \end{align}
where in the last equation we also used the Bianchi identity for $b_{\mu\nu}$. The main difference between these two sets of equations is the factor in front of $\mathfrak{e}^2$ in \eqref{beta3} and \eqref{beta3DFT}. The further difference in equations \eqref{beta5} and \eqref{beta5DFT} arise precisely because of this factor of 2, since one needs to use the scalar equations to prove that the remaining equations are $U(1)$ invariant. In other words, the form of Einstein's equations depends on the factor in front of $\mathfrak{e}^2$ in the scalar equation, because of the requirement of $U(1)$-mass invariance, also see upcoming equations \eqref{lambda1} and \eqref{lambda2}. 
Hence it looks like the DFT and beta functions computations produce almost the same result with the difference being a factor of two in \eqref{beta3} and \eqref{beta3DFT}. It is worth noting that this difference goes away when we consider torsionless geometries, since in that case we have $\mathfrak{e}_\mu =0$ and world-sheet beta functions and double field theory give rise to the same set of equations in the target space.

Despite this difference, both sets of equations describe spacetimes with the same symmetries. In fact, we could generalize them further by modifying equations \eqref{beta3DFT} and \eqref{beta5DFT} by introducing an arbitrary constant $\lambda$:
\begin{align}\label{lambda1}
  D\cdot a + a^2 &=  (1+\lambda) \mathfrak{e}^2 + 2 \left(a \cdot D\phi \right) \, ,\\
  R_{(\mu\nu)} - \frac{1}{4}H^{\rho\sigma}_{\ \ (\mu} H_{\nu)\rho\sigma} + 2 D_{(\mu} D_{\nu)} \phi &= \frac{\mathfrak{e}^2 \left(2\Phi \tau_\mu \tau_\nu-\lambda\, \bar h_{\mu\nu}\right) - \mathfrak{e}_\mu \mathfrak{e}_\nu}{2}-\frac{2a^2\,\Phi \tau_\mu\tau_\nu-a_\mu a_\nu }{2}      \nonumber \\
  &\hphantom{=} +\left(\Delta_S \right)^\rho_{(\mu} D_{\nu)}a_{\rho}  - \mathfrak{e}^\sigma \left( \Delta_T \right)^\rho_{(\mu} H_{\nu)\rho\sigma} \label{lambda2}
\end{align}
while keeping the other equations unchanged. These equations are $U(1)$ invariant for any choice of the constant $\lambda$ and they are written in terms of manifestly boost-invariant quantities. Therefore, they describe theories with extended Galilean symmetry. The only difference between the theories with $\lambda=0$ and $\lambda=1$ is that the former can be found by a variational principle starting from the action $\eqref{TNCaction}$, while the latter can be found by requiring vanishing of Weyl anomaly in the string embedding.

Finally, note that to arrive at the parametrization \eqref{TNCDFT1}-\eqref{TNCDFT2} one needs to perform field redefinitions involving both the $B$-field and $\aleph$. While such field redefinitions are allowed at the classical level, it is not clear whether this will produce the same actions at the quantum level as the path integral measure may, in principle, transform as well. This question is beyond the scope of this work and should be addressed separately in future.  
\subsection{SNC beta functions}
In  \cite{Gomis:2019zyu, YanYu}  the beta functions for a worldsheet with SNC target space have been computed. They imposed the so called \textit{foliation constraint} through their computation:
\begin{equation} \label{FC}
\nabla_{[\mu}\tau_{\nu]}^A=0=F_{\mu\nu}^A.
\end{equation}
If we impose this geometrical constraint, the equations \eqref{FssSNC}-\eqref{QttSNC} become 
\begin{align}\label{SNCFC}
D^\rho H_{\rho}^{\ \mu\nu}  - 2D^\rho \phi \, H_{\rho}^{\ \mu\nu}  &= 0 ,\nonumber\\
D^\mu D_\mu \phi -2 D^\mu \phi \, D_\mu \phi  &= -\frac{1}{12} H^{\rho\mu\nu}H_{\rho\mu\nu},\nonumber\\
\mathcal{R}^{(\mu\nu)} + 2 D^{(\mu}D^{\nu)}\phi -\frac{1}{4}H^\mu_{\ \rho\sigma}H^{\nu \rho\sigma} &=0,\\
h^{\mu \rho}u^{\sigma A}\mathcal{R}_{(\rho\sigma)} +   2 u^{\rho A}D^{\mu}D_{\rho}\phi -\frac{1}{4}H^\mu_{\ \rho\sigma}H_{\lambda}^{\  \rho\sigma}u^{\lambda A}& = \frac{1}{2} \epsilon^{A}_{\ B}u^{\sigma B}\left(D^\rho H^\mu_{\ \rho\sigma} -2u^{\rho B}H^{\mu}_{\ \rho \sigma}D^\rho \phi \right),\nonumber\\
u^{\mu A}u^\nu_{\ A}\mathcal{R}_{\mu\nu}+ 2 u^{\mu A}u^\nu_{\ A}D_\mu D_\nu \phi - \frac{1}{4}u^{\mu A}u^\nu_{\ A}H_{\mu\rho\sigma}H_\nu^{\ \rho\sigma} &=  -\frac{1}{2}\epsilon^{AB}u^{\rho}_{\ A} u^\sigma_{\ B} \left( D^\lambda H_{\rho\sigma \lambda}-2 H_{\rho\sigma \lambda} D^\lambda \phi \right)\nonumber
\end{align}   
where in the first equation we used the fact  that the foliation constraint implies\footnote{Note that this is a sort of "twistlelss condition" on the field strength of $\omega_\mu$.}
\begin{equation}
\Omega_{\mu\nu}=u^\rho_A \tau_{[\mu}^A \Omega_{\nu]\rho}\qquad \Longrightarrow \qquad \Omega^{\mu\nu}=0. 
\end{equation}
Note    that the connection used in the present work is in general different from the one used in the literature, however they are actually equal when the foliation constraint is imposed.  The  connection used in  \cite{Gomis:2019zyu, YanYu} is
\begin{equation}
\bar \Gamma^\rho_{\mu\nu} =\frac{1}{2}h^{\rho\sigma}\left(\partial_{\mu} \bar h_{\nu\sigma}+\partial_{\nu} \bar h_{\mu\sigma}-\partial_{\sigma} \bar h_{\mu\nu}\right)+\frac{1}{2}u^{\rho A}u^\sigma_A\left(\partial_{\mu} \tau_{\nu\sigma}+\partial_{\nu} \tau_{\mu\sigma}-\partial_{\sigma} \tau_{\mu\nu}\right)
\end{equation}
where $\tau_{\mu\nu}\equiv \tau_\mu^A\tau_{\nu A}$. One can check that when $F_{\mu\nu}^A=0$ we have 
\begin{equation}
\Gamma^\rho_{\mu\nu}= \bar \Gamma^\rho_{\mu\nu}.
\end{equation}
This means that the equations \eqref{SNCFC} are exactly the same as the ones that are obtained by requiring the cancellation of the Weyl anomaly in the worldsheet theory.

\subsection{Comparison of TNC and SNC equations of motion}

It was found in  \cite{Harmark:2019upf} that, under certain assumptions which we review below, the SNC worldsheet action reduces to that of TNC. It is then natural to ask  whether the equations of motion of SNC also reduce to the ones of TNC. The basic condition under which SNC reduces to TNC is the presence of an isometry along a compact longitudinal direction. We can then split the SNC spacetime directions as $m=\left(\mu, u\right)$, where $u$ is the compact direction and $\mu$ will describe the TNC directions. Then we impose the following gauge choice on the SNC fields:
\begin{align}
m_M^A&=0, & \tau_\nu^0&=0, &\tau^1_u &=1,& E_u^{A'}=0.
\end{align}
Substituting this ansatz in the equations of motion would be quite tedious. Luckily, since the equations of motion are obtained from an action, which in turn is obtained from the generalized metric of DFT, it will suffice to compare the generalized metrics rather than the equations or the actions. Comparing \eqref{TNCDFT1}-\eqref{TNCDFT2} with \eqref{SNCDFT1} we see that the two parametrizations are indeed the same once we impose the following identifications:
\begin{align}
\upsilon^\mu_1&=0, & \upsilon_1^u&=1,\nonumber\\
\upsilon_0^\mu &= \upsilon^\mu, &\upsilon_0^u &= \upsilon^\mu \aleph_\mu,\nonumber\\
h^{\mu\nu}_{(SNC)}&=h^{\mu\nu}_{(TNC)}, & h^{\mu u}_{(SNC)}&= h^{\mu\rho}\aleph_\rho,\\
h^{uu}_{(SNC)} &= h^{\rho\sigma}\aleph_\rho\aleph_\sigma, & B^{(SNC)}_{\mu\nu}&=B^{(TNC)}_{\mu\nu}\nonumber\\
B^{(SNC)}_{\mu u}&= -m_{\mu}\nonumber. 
\end{align}
In addition, it is not hard to check that the invariant measure of SNC \eqref{SNCmeasure} correctly reduces to the one of TNC \eqref{TNCmeasure}.

This identification implies that the SNC equations of motion  will match the ones obtained for TNC, at least in the DFT formulation. Unfortunately, the  SNC beta functions have only been computed under the assumption that the foliation constraint is satisfied\footnote{The SNC beta functions without the requirement of the foliation constraint were also computed, but only at the linearized level.} , and similarily the   TNC beta functions have been computed assuming twistless torsion, so we cannot compare the full SNC and TNC equations obtained from DFT with the ones obtained by setting the beta functions to zero. However we already showed that when $F_{\mu\nu}^A=0$ the SNC equations \eqref{SNCFC} do indeed match the ones computed in  \cite{Gomis:2019zyu, YanYu}. Using the identification between the SNC and TNC generalized metrics this also implies that the SNC beta functions with $F_{\mu\nu}^A=0$ match the TNC beta functions with $F_{\mu\nu}=0$.

Finally, we have not found any existing results in the literature on actions of Carrollian geometry to compare with our result in section \ref{CarSec}. 

\section{Discussion and Outlook} \label{Conclusions}

In this work we studied a family of non-relativistic gravity theories in the framework of Double Field Theory. In particular we studied three spacetimes which are related to each other: Type I Torsional Newton Cartan, String Newton-Cartan and Carroll. The non-relativistic actions we found are respectively given by equations \eqref{TNCaction}, \eqref{SNCaction} and \eqref{Carrollaction} . As far as we know, these actions are explicitly presented for the first time in the literature. They should correspond to consistent truncations of the related Supergravity actions obtained from string theory and hence are fundamental ingredients in the study of non-relativistic quantum gravity.

These three geometries correspond to a DFT embedding with $n=1=\bar n$, so it is natural to ask how they are related to each other. Relation between Carrollian and TNC geometries was already proposed in  \cite{Hartong:2015xda, Duval:2014uoa}. We found that this connection becomes more clear in the context of DFT: to relate TNC to Carroll we simply have to perform a T-duality on the generalized metric. This is quite a curious result by itself, in that, it implies T-duality in this case  transforms a non-relativistic geometry to the null\footnote{In some sense, ultra-relativistic,  as it is obtained by the limit $c\to 0$.} Carrolian geometry. It would be interesting to figure out physical implications of this. Moreover, we also explicitly showed how one can recover the TNC generalized metric from the SNC one by using the prescriptions proposed in  \cite{Harmark:2019upf}. This means that the three non-relativistic geometries, Torsional Newton-Cartan, Carrollian, and String Newton-Cartan are all related to each other by either a T-duality or a null reduction/uplift.

By varying the actions we also obtained a set of equations of motion for the three aforementioned non-relativistic geometries. In particular we consistently found that each set of equations misses a projected component. As we discussed,  this is a general property of DFT:   $n\times \bar n$ number of equations --- which equals one equation in our case --- would be missing in the final set of equations. The full DFT framework  \cite{Morand:2017fnv} resolves this issue as \eqref{DFTeqs} provides the complete set. This means that theories with $n\times \bar n\neq0$ are to be considered as particular sectors of Double Field Theories rather than independent theories by themselves. When working at the level of low energy effective actions one should then append the missing constraints to the equations of motion derived from the actions \eqref{TNCaction}, \eqref{Carrollaction} and \eqref{SNCaction}. For example, to define a TNC theory we need to provide the action \eqref{TNCaction} \textit{together with} an extra equation, Newton's law, which should be imposed on-shell. This resembles the self-duality condition of the 5-form field in the effective action  for Type IIB String Theory, albeit the origin of the issue is clearly different here.

Once equations of motion were found, we asked whether these equations match the ones already known in the literature. For TNC we showed that the equations in the present work  match the ones obtained from world-sheet beta functions  \cite{Gallegos:2019icg} when $\mathfrak{e} \cdot d \aleph$, that is, contraction of the electric field with the field strength of the Kalb Ramond vector, in (\ref{FTNC}) vanishes. In its presence the difference between the two sets of equations turns out to be merely a factor of $2$. The origin of this discrepancy is not clear to us, see section \ref{TNCbeta} for a more detailed discussion. On the other hand, we showed that the SNC equations of motion obtained from DFT match \textit{precisely} the ones obtained in  \cite{Gomis:2019zyu, YanYu} from the world-sheet beta functions when the foliation constraint is imposed. As already mentioned, the generalized SNC metric reduces to the TNC one in a particular limit, which implies that the equations of motion will also reduce to the TNC ones. Recalling that the foliation constraint reduces to the torsionless limit on the TNC side we conclude that the SNC beta functions  \cite{Gomis:2019zyu, YanYu} reduce to the TNC ones  \cite{Gallegos:2019icg}, which is consistent with what we found by embedding both theories in the DFT framework. It is worth stressing that the discrepancy between the TNC equations and beta functions disappears in the absence of $d \aleph$, which is also implied by the foliation constraint on the SNC side. The full (non-linearized) SNC beta functions with $\nabla_{[\mu}\tau_{\nu]}^A\neq 0$ are not known, therefore we are unable to tell if the discrepancy we find in TNC theory will also appear for SNC.  

The present work can be extended in a number of ways. The most obvious direction is studying solutions to the non-relativistic actions  \eqref{TNCaction}, \eqref{Carrollaction} and \eqref{SNCaction}. Of particular interest would be solutions that are equivalent to black holes in general relativity. The definition of a black hole in a non-Riemannian manifold is unclear  \cite{Bergshoeff:2016soe, Grumiller:2017sjh, Grumiller:2020elf, Arvanitakis:2016zes}, and we believe that the actions presented in this work would help clarify their physical interpretation. In particular, variations of on-shell actions (note that the DFT on-shell action is trivially zero up to boundary terms) would allow for investigation of the thermodynamical properties of such hypothetical black holes and possibly relate them to holographically dual theories in non-relativistic quantum plasmas. Another interesting direction would be to study the relation between TNC and Carroll \eqref{TNCarrDual1}-\eqref{TNCarrDual2}, i.e. understanding in detail the T-duality map between a non-relativistic limit and an ultra-relativistic one.

\begin{acknowledgements}
We wish to thank Eric Bergshoeff, Chris Blair, Troels Harmark, Jelle Hartong,  Johannes Lahnsteiner, Kevin Morand,  Niels Obers,  Gerben Oling,  Jeong-Hyuck Park, Luca Romano, Jan Rosseel, Ceyda \c{S}im\c{s}ek and Ziqi Yan for useful comments and discussions. DG is supported in part by CONACyT through the program Fomento, Desarrollo y Vinculacion de Recursos Humanos de Alto Nivel. This work is partially supported by the Delta-Institute for Theoretical Physics (D-ITP) funded by the Dutch Ministry of Education, Culture and Science (OCW). 
\end{acknowledgements}

\newpage 
\appendix

\section{TNC identities}\label{AppTNC} 

\subsection{Identities}
The connection we use is 
\begin{equation}
\Gamma^\rho_{\mu\nu}= -\hat \upsilon^\rho\partial_\mu \tau_\nu +\frac{1}{2}h^{\rho\sigma}\left(\partial_\mu \bar h_{\nu\sigma}+\partial_\nu \bar h_{\mu\sigma}-\partial_\sigma \bar h_{\mu\nu}\right).
\end{equation}
Integration by parts is not as straightforward as in usual general relativity. Instead we have
\begin{equation}
D_\mu A^\mu = e^{-1}\partial_\mu\left(e\, A^\mu \right)-a_\mu A^\mu +2A^\mu D_\mu \phi,
\end{equation}
where $A^\mu$ is an arbitrary vector.

Many geometric identities can be derived from the completeness relation
\begin{equation}
-\hat\upsilon^\mu\tau_\nu +h^{\mu \rho} \bar h_{\rho\nu}=\delta^\mu_\nu.
\end{equation}
For example we can take the derivative of this relation and then multiply with $\bar h$. This gives
\begin{equation}
D_\rho \bar h _{\mu\nu}=2 \tau_{(\mu}\bar h_{\nu)\lambda}D_\rho \hat\upsilon^\lambda -2\tau_\mu\tau_\nu D_\rho \Phi
\end{equation}
and
\begin{equation}
D_\sigma D_\rho \bar h _{\mu\nu}=2 \tau_{(\mu}\bar h_{\nu)\lambda}D_\sigma D_\rho \hat\upsilon^\lambda +2 \tau_{(\mu}\bar D_\sigma h_{\nu)\lambda} D_\rho \hat\upsilon^\lambda -2\tau_\mu\tau_\nu D_\sigma D_\rho \Phi.
\end{equation}
More useful identities can be derived by the definition of the connection:
\begin{align}
	\begin{split}
		h^{\rho\sigma}D_{\sigma}\bar h_{\mu\nu} &= 4h^{\rho\sigma}F_{\sigma (\mu}\tau_{\nu)} \Phi +2\tau_{(\mu}D_{\nu)}\hat \upsilon^\rho\\
		\hat \upsilon^\rho D_\rho \upsilon^\mu&= h^{\rho\sigma}\left(D_\sigma \Phi +2a_\sigma \Phi \right)\\
		h^{\mu\rho}D_{\rho}\hat \upsilon^\nu&=h^{\nu\rho}D_{\rho}\hat \upsilon^\mu +2\Phi F_{\rho \sigma} h^{\mu \rho} h^{\nu \sigma} .
	\end{split}
\end{align}

\subsection{Variational calculus}
We choose the independent geometric fields to be $h^{\mu\nu},\hat \upsilon^\mu$ and $\Phi$. The variations of the dependant fields are given by
\begin{align} 
	\begin{split}
		\delta \bar h_{\mu\nu}&=-2\tau_\mu\tau_\nu \delta \Phi +2\tau_{(\mu}\bar h_{\nu)\rho}\delta \hat \upsilon^\rho -\bar h_{\mu\rho}\bar h_{\nu\sigma}\delta h^{\rho\sigma}\\
		\delta \tau_{\mu}&=\tau_\mu \tau_{\rho }\delta\hat\upsilon^{\rho} -\bar h_{\mu \rho }\tau_\sigma \delta h^{\rho\sigma}.
	\end{split}
\end{align}
The variation of the measure is then
\begin{equation}
\delta e = \frac{e}{2}\left(\bar h^{\rho\sigma}\delta \bar h _{\rho\sigma}-\frac{\delta \Phi}{\Phi} -4\delta \phi \right) =  \frac{e}{2}\left( -\bar h_{\rho\sigma}\delta h ^{\rho\sigma}+2\tau_\rho\delta\hat\upsilon^\rho-4\delta \phi \right) 
\end{equation}
The variation of the acceleration is
\begin{equation}
\delta a_\mu =  \delta\hat \upsilon^\rho F_{\rho\mu}+2\hat\upsilon^\rho D_{[\rho}\delta \tau_{\mu]}-a_\mu\hat \upsilon^\rho \delta   \tau_\rho. 
\end{equation}
The variation of the connection is
\begin{align}
	\begin{split}
		\delta \Gamma^\rho_{\mu\nu}=&-\frac{1}{2}\delta\hat\upsilon^\rho F_{\mu\nu}-\hat \upsilon D_\mu \delta \tau_\nu -\frac{1}{2}\hat\upsilon F_{\mu\nu} \tau_{\sigma}\delta \hat\upsilon^{\sigma} +\frac{1}{2}h^{\rho\sigma}\left(D_\mu \delta\bar h_{\nu\sigma}+D_\nu \delta\bar h_{\mu\sigma}-D_\sigma \delta\bar h_{\mu\nu}  \right)\\
		& + \frac{1}{2}\delta h^{\rho\sigma}\left(D_\mu \bar h_{\nu\sigma}+D_\nu \bar h_{\mu\sigma}-D_\sigma \bar h_{\mu\nu}  \right) -2\Phi \tau_{(\mu}F_{\nu)\sigma} \delta h^{\rho\sigma} -2\tau_{(\mu}F_{\nu)\sigma} h^{\rho\sigma}\delta \Phi\\
		& -\delta\tau_{(\mu}F_{\nu)\sigma}  h^{\rho\sigma} \Phi +h^{\rho\sigma}\bar h_{\lambda (\mu}F_{\nu)\sigma}\delta \hat\upsilon^\lambda.
			\end{split}
\end{align}
The Palatini identity in the presence of torsion is
\begin{equation}
\delta \mathcal{R}_{\mu\nu}=D_{\rho}\delta \Gamma^\rho_{\nu\mu}-D_{\nu}\delta \Gamma^\rho_{\rho\mu}-2\Gamma^{\sigma}_{[\nu\rho]}\delta \Gamma^\rho_{\sigma\mu}=D_{\rho}\delta \Gamma^\rho_{\nu\mu}-D_{\nu}\delta \Gamma^\rho_{\rho\mu}+\hat\upsilon^\sigma F_{\nu\rho }\delta \Gamma^\rho_{\sigma\mu}.
\end{equation}

The indipendent matter fields are $\aleph_\mu, B_{\mu\nu}$ and $\phi$. The variation of $b_{\mu\nu}$  is
\begin{equation} \label{Varb}
\delta b_{\mu\nu} = 2D_{[\mu}\delta \aleph_{\nu]}-2\Gamma^\rho_{[\mu\nu]}\delta \aleph_\rho
\end{equation}
so that the variation of the electric field is
\begin{align}	\label{Vare}
\delta \mathfrak{e}_\mu = \delta \hat \upsilon^\rho b_{\rho\mu}+ 2D_{[\mu}\delta \aleph_{\nu]}-2\hat \upsilon^\nu\Gamma^\rho_{[\mu\nu]}\delta \aleph_\rho,
\end{align}
while that of  $H_{\mu\nu\rho}$ can be written as
\begin{equation}\label{VarH}
\delta H_{\rho\mu\nu} =3D_{[\rho}\delta B_{\mu\nu]} -2 \Gamma^\sigma_{[\rho\mu]}\delta B_{ \nu\sigma}-2 \Gamma^\sigma_{[\nu\rho]}\delta B_{ \mu\sigma}-2 \Gamma^\sigma_{[\mu\nu]}\delta B_{\rho \sigma}.
\end{equation}
The Kalb-Ramond matter satisfies the usual Bianchi identities
\begin{equation}
dH = db =0.
\end{equation}

In the \textit{twistless} case we have $b_{\mu\nu}=\mathfrak e_\mu \tau_\nu-\mathfrak e_\nu \tau_\mu$, then we can express the Bianchi identity in terms of $\mathfrak e_\mu $ rather than $b_{\mu\nu}$. We find
\begin{equation}
\partial_{[\rho}b_{\mu\nu]}=2D_{[\mu}\mathfrak e_\nu \tau_{\rho]} +2a_{[\mu}\mathfrak e_{\nu}\tau_{\rho]} =0.
\end{equation}
A similar identity can be derived for the acceleration, yielding
\begin{equation}
\partial_{[\rho}F_{\mu\nu]}=2D_{[\mu}a_\nu \tau_{\rho]}  =0.
\end{equation}

\section{Carroll identities} \label{AppCar}

\subsection{Identities}
The connection we utilize when varying the action is
\begin{equation}
\Gamma^\rho_{\mu\nu}= -  \upsilon^\rho\partial_\mu \hat \tau_\nu +\frac{1}{2}\hat h^{\rho\sigma}\left(\partial_\mu   h_{\nu\sigma}+\partial_\nu  h_{\mu\sigma}-\partial_\sigma   h_{\mu\nu}\right),
\end{equation}
from which we find the following identitiy
\begin{equation}
D_\mu A^\mu = e^{-1}\partial_\mu\left(e\, A^\mu \right)-a_\mu A^\mu +2A^\mu D_\mu \phi,
\end{equation} 
valid for any arbitrary vector $A^\mu$.
From the completeness relation and the definition of the connection it is possible to derive the following identities:
\begin{align}
\begin{split}
D_\rho h_{\mu\nu} &= -\mathcal{K}_{\rho\mu}\hat \tau_\nu  -\mathcal{K}_{\rho\nu}\hat \tau_\mu \\
D_\mu \upsilon^\nu &= -\hat h^{\nu\rho}\mathcal{K}_{\rho\mu},\\
\upsilon^\rho D_\rho \upsilon^\mu &=0,
\end{split}
\end{align}
where the extrinsic curvature was defined in the main text as
\begin{equation}  
\mathcal{K}_{\mu\nu}=-\frac{1}{2}\mathcal{L} _{\upsilon} h_{\mu\nu}=-\frac{1}{2}\left( \upsilon^\rho \partial_\rho h_{\mu\nu}+\left(\partial_\mu \upsilon^\rho\right)h_{\rho\nu}+\left(\partial_\nu \upsilon^\rho\right)h_{\mu\rho}\right).
\end{equation}

\subsection{Variational calculus}
We will use $\hat h^{\mu\nu}, \upsilon^\mu, \Phi$ as indipendent geometric fields. The other geometric fields are related to them via
\begin{align} 
	\begin{split}
		\delta   h_{\mu\nu}&=  2\tau_{(\mu}  h_{\nu)\rho}\delta \upsilon^\rho -  h_{\mu\rho} h_{\nu\sigma}\delta \hat h^{\rho\sigma}\\
		\delta\hat \tau_{\mu}&=\tau_\mu \tau_{\rho }\delta \upsilon^{\rho}  - h_{\mu \rho }\tau_\sigma \delta\hat  h^{\rho\sigma}\\
		\delta  \bar h^{\mu\nu}&=  \delta \hat h^{\mu\nu} - 4\Phi \upsilon^{(\mu}\delta \upsilon^{\nu)}-2\delta \Phi \, \upsilon^\mu \upsilon^\nu.
	\end{split}
\end{align}
The variation of the measure is then
\begin{equation}
\delta e = \frac{e}{2}\left(\bar h^{\rho\sigma}\delta \bar h _{\rho\sigma}+\frac{\delta \Phi}{\Phi} -4\delta \phi \right) =  \frac{e}{2}\left(-h_{\mu\nu}\delta \hat h^{\mu\nu}+2 \hat \tau_\mu \delta \upsilon^\mu -4\delta \phi \right) .
\end{equation}
The variation of the acceleration is
\begin{equation}
\delta a_\mu =  \delta \upsilon^\rho F_{\rho\mu}+2\upsilon^\rho D_{[\rho}\delta\hat\tau_{\mu]}- a_\mu \upsilon^\rho \delta \hat \tau_\rho. 
\end{equation}
The variation of the connection is
\begin{align}
	\begin{split}
		\delta \Gamma^\rho_{\mu\nu}=&-\frac{1}{2}\delta \upsilon^\rho F_{\mu\nu}-\upsilon D_\mu \delta \hat \tau_\nu -\frac{1}{2} \upsilon^\rho  F_{\mu\nu}\hat \tau_{\sigma}\delta\upsilon^{\sigma} +\frac{1}{2} \hat h^{\rho\sigma}\left(D_\mu \delta h_{\nu\sigma}+D_\nu \delta h_{\mu\sigma}-D_\sigma \delta h_{\mu\nu}  \right)\\
		& + \frac{1}{2}\delta \hat h^{\rho\sigma}\left(D_\mu  h_{\nu\sigma}+D_\nu h_{\mu\sigma}-D_\sigma  h_{\mu\nu}  \right)  +\hat h^{\rho\sigma} h_{\lambda (\mu}F_{\nu)\sigma}\delta \upsilon^\lambda.
			\end{split}
\end{align}
The Palatini identity  is
\begin{equation}
\delta \mathcal{R}_{\mu\nu}=D_{\rho}\delta \Gamma^\rho_{\nu\mu}-D_{\nu}\delta \Gamma^\rho_{\rho\mu}-2\Gamma^{\sigma}_{[\nu\rho]}\delta \Gamma^\rho_{\sigma\mu}=D_{\rho}\delta \Gamma^\rho_{\nu\mu}-D_{\nu}\delta \Gamma^\rho_{\rho\mu}+ \upsilon^\sigma F_{\nu\rho }\delta \Gamma^\rho_{\sigma\mu}.
\end{equation}

The variations of the Kalb-Ramond matter fields are given by \eqref{Varb}-\eqref{VarH}.

\section{SNC identities}\label{AppSNC}

\subsection{Identities}
The connection is 
\begin{equation}
\Gamma^\rho_{\mu\nu}= -u^\rho_A \nabla_\mu \tau_\nu^A +\frac{1}{2}h^{\rho\sigma}\left(D_\mu \bar h_{\nu\sigma}+D_\nu \bar h_{\mu\sigma}-D_\sigma \bar h_{\mu\nu}\right)
\end{equation}
where we introduced the spin connection via
\begin{equation}
\nabla_\mu \tau_\nu^A= \partial_\mu \tau_\nu^A +\Omega_\mu^{\ AB}\tau_{\nu B}\equiv  \partial_\mu \tau_\nu^A +\omega_\mu \epsilon^{A}_{\ B}\tau_{\nu}^{\ B}
\end{equation}
and our convention for the longitudinal epsilon symbol is $\epsilon_{01}=+1$. This connection is boost invariant and compatible with $h^{\mu\nu}$ and $\tau_\mu^A$,
\begin{equation}
D_\rho h^{\mu\nu}=D_\rho \tau_\mu^A=0,
\end{equation}
where we are using the symbol $D$ (not to be confused with the dimensionality of spacetime) to denote the full covariant derivative, i.e.
\begin{equation}
D_\mu \tau_\nu^A = \partial_\mu \tau_\nu^A -\Gamma^\rho_{\mu\nu} \tau_\rho^A +\omega_\mu \epsilon^A_{\ B}\tau_\nu^B=0.
\end{equation}
Integration by parts is then performed with the use of the identity
\begin{equation}
D_\mu A^\mu = e^{-1}\partial_\mu\left(e\, A^\mu \right)-a_\mu A^\mu +2A^\mu D_\mu \phi,
\end{equation}
where we recall $a_\mu=a_\mu^{AB} \eta_{AB}$.

From the completeness relation we find the following decomposition
\begin{equation}
D_\rho \bar h_{\mu\nu} = 2 \tau_{(\mu}^{\ \ A} \bar h_{\nu)\sigma} D_\rho u^{\sigma}_{\ A}-\tau_\mu^A\tau_\nu^B D_\rho \Phi _{AB}.
\end{equation}
 From the definition of the connection we find
\begin{align}
\begin{split}
h^{\rho\sigma}\left(D_\mu \bar h_{\nu\sigma} +D_\nu\bar h_{\mu\sigma}-D_\sigma \bar h_{\mu\nu} \right)&=-2 h^{\rho\sigma}  F_{\sigma (\mu}^{\ \ \ A}\tau_{\nu)}^{\ B} \Phi_{AB} .
\end{split}
\end{align}
From the projections of this identity it follows that
\begin{equation}
h^{\rho[\mu}D_\rho u^{\nu]}_A = \frac{1}{2}F^{\mu \nu B} \Phi_{BA}
\end{equation}
and
\begin{equation}
u^\nu_{(A} D_\nu u^\mu_{B)}=h^{\mu\nu}\left( a_{\nu (A}^{\ \ \ \, C}\Phi_{B)C}   +\frac{1}{2}D_\nu \Phi_{AB}  \right). 
\end{equation}

Since $\Phi_{AB}$ is a $2\times 2$ symmetric matrix, we can write its inverse as
\begin{equation}
(\Phi^{-1})_{AB}= \frac{\Phi_{AB}-\eta_{AB} \Phi}{\det \Phi}
\end{equation}
where $\det \Phi = \frac{1}{2} \epsilon^{AB} \epsilon^{CD}\Phi_{AC}\Phi_{BD}$ and $\Phi=\Phi^A_A$. Moreover since the longitudinal indices can only take two different values, we have that 
\begin{equation}
T^{[\alpha \beta \gamma\dots ]}=0
\end{equation}
for any tensor $T$ with three or more longitudinal indices.

The field strength of $\tau_\mu^A$ can be decomposed as
\begin{equation}
F_{\mu\nu}^{ A}=f^A \tau_{\mu}^B    \tau_{\nu}^C \epsilon_{BC} + 2a_{[\mu}^{B A} \tau_{\nu] B}+\mathcal{F}_{\mu\nu}^{ A}
\end{equation}
with
\begin{equation}
u^{\mu}_A \mathcal{F}_{\mu\rho B}= 0, \qquad \qquad  u^{\mu A}a_{\mu}^{\ BC}=\epsilon^{AB}f^C.
\end{equation}

\subsection{Variational calculus}
 The indipendent fields are $u^\mu_A, h^{\mu\nu}, \Phi^{AB}$ and $B_{\mu\nu}$. The other SNC fields are related via
\begin{align}
	\begin{split}
		\delta \tau_\mu^A &= \tau_\mu^B \tau_\rho^A \delta u^\rho_B   - \tau_\rho^A \bar h_{\mu\sigma} \delta h^{\rho\sigma}\\ 
\delta \bar h_{\mu\nu} &=2\tau_{(\mu}^A\bar h_{\nu)\rho}u^{\rho}_A -\bar  h_{\mu\rho}\bar h_{\nu\sigma}\delta h^{\rho\sigma} -\tau_\mu^A\tau_\nu^B \delta \Phi_{AB}.
	\end{split}
\end{align}
The variation of the connection is
\begin{align}
	\begin{split}
		\delta \Gamma^\rho_{\mu\nu} &=-u^\rho_A \left(D _\mu \delta \tau_\nu^A +\Gamma^\sigma_{[\mu\nu]}\delta \tau_\sigma^A + \epsilon^A_{\ B}\tau_{\nu B}\delta \omega_\mu\right)  -  \delta u^\rho_A \Gamma^\sigma_{[\mu\nu]} \tau_\sigma^{A} \\
&\hphantom{=} +\frac{1}{2}h^{\rho\sigma}\left(D_\mu \delta \bar h_{\nu\sigma}+D_\nu \delta \bar h_{\mu\sigma} -D_\sigma \delta \bar h_{\mu\nu}\right)+\frac{1}{2}\delta h^{\rho\sigma}\left(D_\mu \bar h_{\nu\sigma}+D_\nu\bar  h_{\mu\sigma} -D_\sigma \bar h_{\mu\nu}\right)\\
&\hphantom{=}+h^{\rho\sigma}\left(\Gamma^{\lambda}_{[\mu\sigma]}\delta \bar h_{\lambda \nu} +\Gamma^{\lambda}_{[\nu\sigma]}\delta \bar h_{\lambda \mu}\right)+\delta h^{\rho\sigma}\left(\Gamma^{\lambda}_{[\mu\sigma]} \bar h_{\lambda \nu} +\Gamma^{\lambda}_{[\nu\sigma]} \bar h_{\lambda \mu}\right).
	\end{split}
\end{align}
The variation of the Ricci scalar can be found as usual through the Palatini identity
\begin{equation}
\delta R_{\mu\nu}=D_\rho \delta \Gamma^\rho_{\nu\mu}-D_\nu \delta \Gamma^\rho_{\rho\mu} -2\Gamma^\rho_{[\nu\sigma]}\delta \Gamma^\sigma_{\rho\mu}=D_\rho \delta \Gamma^\rho_{\nu\mu}-D_\nu \delta \Gamma^\rho_{\rho\mu} -2u^\rho_A\nabla_{[\nu}\tau_{\sigma]}^A\delta \Gamma^\sigma_{\rho\mu}.
\end{equation}

The variation of the field strength of the $B$-field  can be read off from   \eqref{VarH}.

\section{Actions in Einstein frame}
In this section we will perform a conformal redefinition of the metric complex to rewrite the non-relativistic string frame actions \eqref{TNCaction}, \eqref{Carrollaction}, \eqref{Carrollaction2} and \eqref{SNCaction} in Einstein frame.

\subsection{TNC}
The action and equations of motion can be written in terms of the basic fields $h^{\mu\nu}$ , $\hat{v}^{\mu}$ and $\Phi$.
The aforementioned fields transform as
\begin{align}
\begin{split}
    h^{\mu \nu} &\rightarrow e^{-\alpha \phi} h^{\mu \nu},\\
    \upsilon^{\mu} &\rightarrow e^{-\alpha \phi} \upsilon^{\mu},\\
    \Phi &\rightarrow e^{-\alpha \phi} \Phi,
\end{split}
\end{align}
where $e^{-\alpha \phi}$ is the conformal factor and $\alpha = \frac{4}{d-1}$.  Then, in Einstein frame, the TNC action \eqref{TNCaction} can be rewritten as
\begin{align}
	\begin{split}
		S =& \int d^d x\, \sqrt{\frac{\det \bar h_{\mu\nu}}{2\Phi}}\, \left[ \mathcal{R}+ \frac{1}{2}a^\mu a_\mu + \frac{1}{2} e^{-2\alpha \phi}\mathfrak{e}^\mu \mathfrak{e}_\mu - \alpha D^\mu\phi D_\mu \phi   - \frac{1}{12} e^{-2\alpha\phi} H^{\mu \nu\rho}H_{\mu\nu\rho}\right.\\
		&\hphantom{((( \int d^d x\, \sqrt{\frac{\det \bar h_{\mu\nu}}{2\Phi}}}   \left.   - \frac{1}{2} e^{-2\alpha\phi} \hat \upsilon^\rho H_{\rho\mu\nu } b^{\mu\nu} - \frac{1}{2} \left(F^{\mu \nu} F_{\mu \nu}+e^{-2\alpha\phi} b^{\mu\nu}b_{\mu \nu} \right)\Phi  \right]
	\end{split}
\end{align}

\subsection{Carroll}
The action and equations of motion can be written in terms of the basic fields $\hat h^{\mu\nu}$ , ${v}^{\mu}$ and $\Phi$.
The aforementioned fields transform as
\begin{align}
\begin{split}
    \hat h^{\mu \nu} &\rightarrow e^{-\alpha \phi} \hat h^{\mu \nu},\\
    \upsilon^{\mu} &\rightarrow e^{-\alpha \phi} \upsilon^{\mu},\\
    \Phi &\rightarrow e^{+\alpha \phi} \Phi,
\end{split}
\end{align}
where once again we have $\alpha = \frac{4}{d-1}$.  In Einstein frame the action \eqref{Carrollaction} is
\begin{align}
	\begin{split}
		S =& \int d^d x\,  \sqrt{2\Phi \det \bar h_{\mu\nu}} \, \left[ \mathcal{R} + \frac{1}{2}a^\mu a_\mu + \frac{1}{2}e^{-2\alpha \phi}\mathfrak{e}^\mu \mathfrak{e}_\mu - \alpha D^\mu\phi D_\mu \phi +2 \mathcal{K}  \upsilon^\mu D_\mu \Phi    \right.\\
		&\hphantom{(((\int d^d x \sqrt{2\Phi \det \bar h_{\mu\nu}}  }   \left.  + 2\Phi\left(\mathcal{K}^{\mu\nu}\mathcal{K}_{\mu\nu}-\mathcal{K}^2 + \alpha\, \upsilon^\mu D_\mu \phi\,\upsilon^\nu D_\nu \phi +\frac{1}{4}e^{-2\alpha \phi}\upsilon^\rho \upsilon^\sigma H^{\mu\nu}_{\hphantom{\mu\nu}\rho}H_{\mu\nu\sigma}    \right)\right.  \\
		&\hphantom{(((\int d^d x \sqrt{2\Phi \det \bar h_{\mu\nu}}  }   \left. -\frac{1}{12}e^{-2\alpha \phi} H_{\mu\nu\rho}H^{\mu\nu\rho}+\frac{1}{2}e^{-2\alpha \phi}b^{\mu\nu}H_{\mu\nu\rho}\upsilon^\rho   \right]
	\end{split}
\end{align}
Similarily we can transform the action \eqref{Carrollaction2}:
\begin{align}
	\begin{split}
		S =& \int d^d x\, \sqrt{2\Phi \det \bar h_{\mu\nu}}\, \left[ \mathcal{\bar R} + \frac{1}{2}a^\mu a_\mu + \frac{1}{2}e^{-2\alpha \phi}\mathfrak{e}^\mu \mathfrak{e}_\mu-\alpha\,\bar D^\mu\phi \bar D_\mu \phi  \right.\\ 
		&\hphantom{(((\int d^d x \sqrt{2\Phi \det \bar h_{\mu\nu}}  }   \left. -\frac{1}{12}e^{-2\alpha \phi}H_{\mu\nu\rho}H^{\mu\nu\rho}+\frac{1}{2}e^{-2\alpha \phi}b^{\mu\nu}\upsilon^\rho H_{\mu\nu\rho}  \right],
	\end{split}
\end{align}

\subsection{SNC}
The action and equations of motion can be written in terms of the basic fields $ h^{\mu\nu}$ , ${v}^{\mu}_A$ and $\Phi_{AB}$.The aforementioned fields transform as
\begin{align}
\begin{split}
     h^{\mu \nu} &\rightarrow e^{-\alpha \phi} h^{\mu \nu},\\
    u^{\mu}_A &\rightarrow e^{-\alpha \phi} u^{\mu}_A,\\
    \Phi_{AB} &\rightarrow e^{-\alpha \phi} \Phi_{AB},
\end{split}
\end{align}
where  $\alpha = \frac{4}{D}$ for SNC. The action \eqref{SNCaction} can be written in Einstein frame as
\begin{align}
	\begin{split}
		S =& \int d^D x\, \sqrt{\frac{\det \bar h_{\mu\nu}}{\det \Phi_{AB}}}\, \left[ \mathcal{R} -a^{\mu AB} (a_{\mu (AB)}  -\frac{1}{2}\eta_{AB} a_\mu )+a^\mu a_\mu + \alpha\, a^\mu D_\mu \phi +\alpha(\alpha-1) D^\mu \phi\, D_\mu \phi  \right.\\
		&\hphantom{(((\int d^d x e\, }   \left.  -\frac{1}{2} F^{\mu\nu A}F_{\mu\nu}^{B} (\Phi_{AB}-\frac{1}{2} \eta_{AB}\Phi )  +\frac{1}{2}e^{-\alpha \phi}\epsilon_{AB} u^{\rho A}F^{\mu\nu B}H_{\rho\mu\nu}   -\frac{1}{12}e^{-2\alpha \phi}H^{\rho\mu\nu}H_{\rho\mu\nu}    \right]  \\ 
	\end{split}
\end{align}

\newpage

\bibliographystyle{ieeetr} 
\bibliography{nnbib}

\end{document}